\documentclass{aa}

\usepackage{epsfig}
\usepackage{graphics}
\usepackage[latin1]{inputenc}
\usepackage{natbib}

\usepackage{amssymb}
\usepackage{amsmath}
\usepackage{widetext}

\newcommand{\h}{\bar{H}}
\newcommand{\x}{\bar{x}}

\begin{document}
\title {Experimental constraints on the uncoupled Galileon model from SNLS3 data and other cosmological probes}
\author{J.~Neveu\inst{1}, V.~Ruhlmann-Kleider\inst{1}, A.~Conley\inst{2}, 
N.~Palanque-Delabrouille\inst{1},
P.~Astier\inst{3}, J.~Guy\inst{3}, E. Babichev\inst{4,5}
}
\institute{CEA, Centre de Saclay, Irfu/SPP,  91191 Gif-sur-Yvette, France
\and Center for Astrophysics and Space Astronomy, University of Colorado, Boulder, CO 80309-0389, USA
\and  LPNHE, Universit\'e Pierre et Marie Curie, Universit\'e Paris Diderot, CNRS-IN2P3, 4 place Jussieu, 75252 Paris Cedex 05, France
\and Laboratoire de Physique Th\'eorique d'Orsay, B\^atiment 210,
Universit\'e Paris-Sud 11, F-91405 Orsay Cedex, France
\and $\mathcal{G}\mathbb{R}\varepsilon\mathbb{C}\mathcal{O}$, Institut d'Astrophysique de Paris, UMR 7095-CNRS, Universit\'e Pierre et Marie Curie-Paris 6, 98bis boulevard Arago, F-75014 Paris, France
}

\date{\today}
\authorrunning{J. Neveu et al.}
\titlerunning{Experimental constraints on the uncoupled Galileon model}

\abstract{}{The Galileon model is a modified gravity theory that may provide an explanation for the accelerated expansion of the Universe. This model does not suffer from instabilities or ghost problems (normally associated with higher-order derivative theories), restores local General Relativity -- thanks to the Vainshtein screening effect -- and predicts late-time acceleration of the expansion.}{We derive a new definition of the Galileon parameters that allows us to avoid having to choose initial conditions for the Galileon field. We tested this model against precise measurements of the cosmological distances and the rate of growth of cosmic structures.}
{We observe a weak tension between the constraints set by growth data and those from distances. However, we find that the Galileon model remains consistent with current observations and is still competitive with the $\Lambda$CDM model, contrary to what was concluded in recent publications.}{}
\maketitle

\section{Introduction}





The discovery of the accelerated expansion of the Universe \citep{bib:riess,bib:perlmutter} led cosmologists to introduce dark energy to explain our Universe. Adding a cosmological constant ($\Lambda$) to Einstein's General Relativity is the simplest way to interpret observational data. However, even if adding a new fundamental constant is satisfactory, the value of $\Lambda$ obtained from numerous measurements results in significant fine-tuning and coincidence problems. Thus, there is theoretical motivation to find alternative explanations, such as modified gravity models.

The Galileon model is just such a formulation. It was first proposed by \cite{bib:nicolis} as a general theory involving a scalar field, hereafter called $\pi$, and a second-order equation of motion invariant under a Galilean shift symmetry ($\partial_\mu \pi \rightarrow \partial_\mu \pi + b_\mu$, where $b_\mu$ is a constant vector). This symmetry was first noticed in braneworld theories such as the DGP model of \cite{bib:dgp}. 
The DGP model has the advantage of providing a self-accelerating solution to explain the expansion of the Universe, but it is plagued by ghost and instability problems. Galileon theories are a generalization of the DGP model that avoid these problems. The Galileon model was derived in a covariant formalism by \cite{bib:deffayet}. It was also shown that this model forms a subclass of the general tensor-scalar theories involving only up to second-order derivatives originally found by Horndeski \citep{bib:horndeski}.

In a four-dimension spacetime, only five Lagrangian terms are possible when forming
an equation of motion for $\pi$ invariant under the Galilean symmetry. Therefore, the Galileon Lagrangian has only five parameters.
In the Galileon theory, as in the DGP theory, a screening mechanism called the Vainshtein effect \citep{bib:vainshtein} arises
near massive objects due to non-linear derivative self-couplings of the $\pi$ field. These ensure that the Galileon fifth force is screened near massive objects, and preserves General Relativity on local scales where it has been experimentally tested to high precision. However, this screening is only effective below a certain distance from massive objects (the Vainshtein radius) that depends on the mass of the object and on the values of the Galileon parameters \citep{bib:burrage}.
Experimental constraints on the Galileon parameters based on local tests of gravity have been proposed by \citet{bib:brax11} and \cite{bib:babichev}.

Recently, the Galileon model has been tested against observational cosmological data by \citet{bib:appleby2}, \citet{bib:okada}, and \citet{bib:nesseris}. These authors tend to reject the Galileon model because of tensions between growth-of-structure constraints and the other cosmological probes. The evolution of the Universe in the Galileon theory is based on differential equations involving the $\pi$ field, which requires one to set initial conditions, and the above studies resorted to different methods for setting these initial conditions. In this work, we avoid this problem by introducing a new parametrization of the Galileon model that renders it independent of initial conditions. Combined with theoretical constraints derived in \citet{bib:appleby} and \citet{bib:felice2011}, we compare our model with cosmological observables, and find that the Galileon model is not significantly disfavored by current observations.

We used the most recent measurements of type Ia supernovae (SN~Ia) luminosity distances, the cosmic microwave background (CMB), and baryon acoustic oscillations (BAO). The highest-quality SN~Ia sample currently available is the SNLS3 sample described in \cite{bib:guy2010}, \cite{bib:conley}, and \cite{bib:sullivan}. For the CMB, we used the observables from WMAP7 \citep{bib:komatsu11} and the set of BAO distances of the BOSS analysis \citep{bib:sanchez}. The growth of structures is an important probe for distinguishing modified gravity models from standard cosmological models such as $\Lambda$CDM, so it has to be used carefully. 
In this work, we used $f\sigma_8(z)$ measurements from several surveys, corrected for the Alcock-Paczynski effect.

Section~\ref{sec:cosgal} provides the Galileon equations used to compute the evolution of the Universe and the theoretical constraints imposed on the Galileon field. Section~\ref{sec:data} describes the likelihood analysis, data samples, and the computing of cosmological observables. Section~\ref{sec:results} gives the constraints on the Galileon model derived from data,
and Sect.~\ref{sec:disc} discusses these results and their implications. We conclude in Sect.~\ref{sec:concl}.

\section{Cosmology with Galileons}\label{sec:cosgal}

\subsection{Lagrangians}\label{sec:lagrangians}

\par The Galileon model is based on the assumption that the scalar field equation of motion is invariant under Galilean symmetries: $\partial_\mu \pi \rightarrow \partial_\mu\pi + b_\mu$, where $b_\mu$ is a constant four vector. By imposing this symmetry, \cite{bib:nicolis} showed that there are only five possible Lagrangian terms $L_i$ for the Galileon model action. The covariant formulation of the Galileon Lagrangian was derived in \cite{bib:deffayet}. In this paper we start with this covariant action with the parametrization of~\cite{bib:appleby}:

\begin{equation}\label{eq:action}
S=\int d^4x \sqrt{-g}\left( \frac{M_P^2R}{2} - \frac{1}{2} \sum_{i=1}^{5} c_i L_i - L_m \right),
\end{equation}
with $L_m$ the standard-matter Lagrangian, $M_P$ the Planck mass, $R$ the Ricci scalar, and g the determinant of the metric. 
The $c_i$s are the arbitrary dimensionless parameters of the Galileon model that weight the different terms. The Galileon Lagrangians have a covariant formulation derived in \cite{bib:deffayet}:

\begin{widetext}
\begin{equation*}
L_1=M^3\pi,\quad L_2=(\nabla_\mu \pi)(\nabla^\mu \pi),\quad
L_3=(\square \pi)(\nabla_\mu \pi)(\nabla^\mu \pi)/M^3 
\end{equation*}
\begin{equation}
L_4=(\nabla_\mu \pi)(\nabla^\mu \pi)\left[ 2(\square \pi)^2 - 2 \pi_{;\mu\nu}\pi^{;\mu\nu} - R(\nabla_\mu \pi)(\nabla^\mu \pi)/2 \right]/M^6
\end{equation} 
\begin{equation*}
L_5=(\nabla_\mu \pi)(\nabla^\mu \pi)
\left[ (\square \pi)^3 - 3(\square \pi) \pi_{;\mu\nu}\pi^{;\mu\nu}
+2\pi_{;\mu}\ ^{;\nu}\pi_{;\nu}\ ^{;\rho}\pi_{;\rho}\ ^{;\mu} -6 \pi_{;\mu}\pi^{;\mu\nu}\pi^{;\rho}G_{\nu\rho}\right]/M^9,
\end{equation*}
\end{widetext}
\noindent where $M$ is a mass parameter defined as $M^3=H_0^2M_P$, where $H_0$ is the current value of the Hubble parameter.  With this definition the $c_i$s are dimensionless.

$L_2$ is the usual kinetic term for a scalar field, while $L_3$ to $L_5$ are non-linear couplings of the Galileon field to itself, to the Ricci scalar $R$, and to the Einstein tensor $G_{\mu\nu}$, providing the necessary features for modifying gravity and mimicking dark energy. $L_1$ is a tadpole term that acts as the usual cosmological constant, and may furthermore lead to vacuum instability because it is an unbounded potential term. Therefore, in the following we set $c_1=0$.

\cite{bib:appleby} proposed additional direct linear couplings to matter to add to the action: a linear coupling to matter $L_0=c_0 \pi T^\mu_\mu/M_P$ and a derivative coupling to matter $L_G=c_G \partial_\mu \pi \partial_\nu \pi T^{\mu\nu}/(M_P M^3)$, which arises in some brane-world theories (see e.g. \cite{bib:trodden}), where $T^{\mu\nu}$ is the matter energy-momentum tensor. 
These couplings may modify the physical origin of the accelerated expansion of the Universe. Without coupling, the Universe is accelerated only because of the back-reaction of the metric to the energy-momentum tensor of the scalar field, and the Galileon acts as a dark energy component. If the Galileon is coupled directly to matter, instead, it can give rise to accelerated expansion in the Jordan frame, while the Einstein-frame expansion rate is not accelerating. In that case, the cosmic acceleration stems entirely from a genuine modified gravity effect. In this work, we do not consider these optional extensions to the theory, so the Einstein frame and Jordan frame coincide. For more information about the Einstein and Jordan frames, see e.g. \cite{bib:faraoni}.

Action \ref{eq:action} leads to three differential equations: two Einstein equations ((00) temporal component and (ij) spatial component) coming from the variation of the action with respect to the metric $g_{\mu\nu}$, and the scalar field equation of motion from the variation of the action with respect to the $\pi$ field. The equations are given explicitly in Appendix B of \cite{bib:appleby}. With these three differential equations the evolution of the Universe and the dynamics of the field can be computed. 

To solve the cosmological equations, we chose the Friedmann-Lema\^itre-Robertson-Walker (FLRW) metric. With no direct couplings, the functions to compute are the Hubble parameter $H=\dot a /a$ (with $a$ the cosmic scale factor), and $x=\pi'/M_P$, with a prime denoting $d/d \ln a$ (see \cite{bib:appleby} and Sect.~\ref{sec:eq}).

\subsection{Initial conditions}\label{sec:inicond}

\par To compute the solutions of the above equations, we need to set one initial condition for $x$.
We arbitrarily chose to define this initial condition at $z=0$, which we denote $x_0=x(z=0)$. Unfortunately, we have no prior information about the value of the Galileon field or its derivative at any epoch. Fortunately, $x_0$ can be absorbed by redefining the $c_i$s as follows:
\begin{align}
\bar  c_i &= c_i x_0^i\\
\x &= x/x_0\\
\h &= H/H_0 .
\end{align}
This redefinition allows us to avoid treating $x_0$ as an extra free parameter of the model\footnote{If the optional coupling parameters $c_0$ and $c_G$ are included, they should be redefined as $\bar c_0 = c_0 x_0$ and $\bar c_G = c_G x_0^2$. But if $c_0\neq 0$, 
two initial conditions are needed ($\pi_0$ and $\pi'_0$). With our new parametrization we would have to introduce and fit a new parameter $r_0=\pi_0/\pi'_0$.}. Doing so, the $\bar c_i$s remain dimensionless, and the initial conditions are simple:
\begin{equation}
\x_0 = 1,\quad \h_0 = 1 .
\end{equation}

Note that the (00)~Einstein equation could also be used as a constraint equation to fix $x_0$ (see Appendix A) given a set of cosmological parameters $c_i$s, $\Omega_m^0$ and $\Omega_r^0$. If we were to adapt this, we would observe a degeneracy between the parameters: the same cosmological evolution can be obtained with small $c_i$s and a high $x_0$, or with high $c_i$s and a small $x_0$. In other words, different sets of parameters $\left\lbrace c_i,x_0 \right\rbrace$ produce the same cosmology, i.e., the same $\rho_{\pi}(z)$, which is undesirable. Our parametrization avoids this problem by absorbing the degeneracy between the $c_i$s and $x_0$ into our $\bar c_i$s.

\subsection{Cosmological equations}\label{sec:eq}

\par To compute cosmological evolution in the Galileon model, we assume for simplicity that the Universe is spatially flat, in agreement with current observations. We used the Friedmann-Lema\^itre-Robertson-Walker (FLRW) metric in a flat space: 
\begin{equation}
ds^2=-dt^2+a^2\delta_{ij}dx^idx^j .
\end{equation}
When writing the cosmological equations, we can mix the (ij) Einstein equation and the $\pi$ equation of motion to obtain the following system of differential equations for $\x$ and $\h$:

\begin{align}
\x' &=-\x + \frac{\alpha\lambda - \sigma\gamma}{\sigma\beta - \alpha\omega} \label{eq:dpi}\\
\h'  &= \frac{\omega\gamma - \lambda\beta}{\sigma\beta - \alpha\omega} \label{eq:dh}
\end{align}
with
\begin{align}
\alpha &= \frac{\bar  c_2}{6}\h \x -3\bar  c_3\h^3\x^2 + 15\bar  c_4\h^5\x^3  -\frac{35}{2}\bar  c_5\h^7\x^4 \\
\gamma &= \frac{\bar  c_2}{3}\h^2\x - \bar  c_3\h^4\x^2  + \frac{5}{2}\bar  c_5\h^8\x^4 \\
\beta &= \frac{\bar  c_2}{6}\h^2 -2\bar  c_3\h^4\x + 9\bar  c_4\h^6\x^2 - 10\bar  c_5\h^8\x^3 \\
\sigma &= 2\h + 2\bar  c_3\h^3\x^3 - 15\bar  c_4\h^5\x^4 + 21\bar  c_5\h^7\x^5\\
\begin{split}
\lambda &= 3\h^2 + \frac{\Omega_r^0}{a^4} +\frac{\bar c_2}{2}\h^2\x^2 - 2\bar  c_3\h^4\x^3 \\ &\ \ + \frac{15}{2}\bar c_4\h^6\x^4  - 9\bar  c_5 \h^8\x^5 \label{eq:lambda}
\end{split}\\
\omega &= 2\bar  c_3\h^4\x^2 - 12\bar  c_4\h^6\x^3 + 15\bar c_5 \h^8\x^4,
\end{align}
as derived in the formalism of \cite{bib:appleby}, but using our normalization for the $c_i$s. We obtain the same equations except that the $c_i$s are changed into $\bar c_i$s, and that we have a different treatment for the initial conditions. Equations~\ref{eq:dpi} and \ref{eq:dh} depend only on the $\bar c_i$s and $\Omega_r^0$. The radiation energy density in equation \ref{eq:lambda} is computed from the usual formula $\Omega_r^0=\Omega_\gamma^0(1+0.2271 N_{\mathrm{eff}})$ with $N_{\mathrm{eff}}=3.04$ the standard effective number of neutrino species \citep{bib:mangano}.  The photon energy density at the current epoch is given by
$\Omega_\gamma^0h^2=2.469\times 10^{-5}$ (where, as usual, $h=H_0/(100$ km/s/Mpc) for $T_{\mathrm{CMB}} =2.725\ \mathrm{K}$. 

\subsection{Perturbation equations}\label{sec:pert}

\par To test the Galileon model predictions for the growth of structures, we also need the equations describing density perturbations. We followed the approach of \citet{bib:appleby} for the scalar perturbation. \citet{bib:appleby} performed their computation in the frame of the Newtonian gauge, for scalar modes in the subhorizon limit, with the following perturbed metric:

\begin{equation}
ds^2=-(1+2\psi)dt^2+a^2(1-2\phi)\delta_{ij}dx^idx^j.
\end{equation}

In this context, the perturbed equations of the (00)~Einstein equation, the (ij) Einstein equation, the $\pi$ equation of motion, and the equation of state of matter are in the quasi-static approximation

\begin{align}
\frac{1}{2} \kappa_4 \bar \nabla^2 \psi - \kappa_3\bar \nabla^2 \phi &= \kappa_1\bar \nabla^2 \delta y \label{eq:perturbed1}\\
\kappa_5 \bar \nabla^2 \delta y - \kappa_4 \bar \nabla^2 \phi &= \frac{a^2 \rho_m}{H_0^2 M_P^2}\delta_m \\
\frac{1}{2} \kappa_5 \bar \nabla^2 \psi - \kappa_1\bar \nabla^2 \phi &= \kappa_6\bar \nabla^2 \delta y \\
\h^2\delta_m''+\h\h'\delta_m'+&2\h^2 \delta_m'=\frac{1}{a^2}\bar \nabla^2 \psi, \label{eq:perturbed2}
\end{align}
where $\delta y = \delta \pi / M_P$ is the perturbed Galileon, $\bar \nabla = \nabla / H_0$, $\rho_m$ is the matter density, and $\delta_m=\delta \rho_m / \rho_m$ is the contrast matter density. $\kappa_i$s are the same as in \citet{bib:appleby}, but rewritten following our parametrization:
\begin{align}
\begin{split}
\kappa_1 &= -6\bar c_4 \h^3\x^3\left(\h'\x+\h\x'+\frac{\h\x}{3}\right) \\ &\ \ +\bar c_5\h^5\x^3(12\h\x' + 15\h'\x + 3\h\x) 
\end{split}\\
\kappa_3 &= -1 -\frac{\bar c_4}{2}\h^4\x^4 - 3\bar c_5 \h^5\x^4(\h'\x+\h\x') \\
\kappa_4 &= -2 + 3 \bar c_4 \h^4\x^4 - 6 \bar c_5 \h^6\x^5\\
\kappa_5 &= 2 \bar c_3 \h^2\x^2 - 12\bar c_4 \h^4 \x^3 + 15 \bar c_5 \h^6\x^5\\
\begin{split}
\kappa_6 &= \frac{\bar c_2}{2} - 2\bar c_3(\h^2\x' + \h\h'\x + 2\h^2\x) \\ &\ \ + \bar c_4 (12 \h^4\x\x' + 18 \h^3\x^2\h' + 13 \h^4\x^2) \\ &\ \ - \bar c_5(18\h^6\x^2\x'+30 \h^5\x^3\h' + 12 \h^6\x^3) .
\end{split}
\end{align}

With equations \ref{eq:perturbed1} to \ref{eq:perturbed2}, we can obtain a Poisson equation for $\psi$, with an effective gravitational coupling $G_{\mathrm{eff}}^{(\psi)}$ that varies with time and depends on the Galileon model parameters $\bar c_i$s:
\begin{equation}
\bar \nabla^2 \psi = \frac{4\pi a^2 G_{\mathrm{eff}}^{(\psi)}\rho_m}{H_0^2}\delta_m
\end{equation}
\begin{equation}\label{eq:geff}
G_{\mathrm{eff}}^{(\psi)}=\frac{4(\kappa_3 \kappa_6 - \kappa_1^2) }{\kappa_5(\kappa_4 \kappa_1 - \kappa_5 \kappa_3) - \kappa_4(\kappa_4 \kappa_6 - \kappa_5 \kappa_1)} G_N,
\end{equation}
with $G_N$ Newton's gravitational constant.  These equations can be used to compute the growth of matter perturbations in the frame of the Galileon model (see Sect.~\ref{sec:gof}). Tensorial perturbations modes also exist, and are studied in Sect.~\ref{sec:tensorial}.

\subsection{Theoretical constraints}\label{sec:theoconstraints}

\par With so many parameters, it is necessary to restrict the parameter space theoretically before comparing the model to data. The theoretical constraints arise from multiple considerations: the (00)~Einstein equation, requiring positive energy densities, and avoiding instabilities in scalar and tensorial perturbations.

\subsubsection{The (00)~Einstein equation and $\bar c_5$}

\par Because we used only the (ij) Einstein equation and the $\pi$ equation of motion to compute the dynamics of the Universe (equations \ref{eq:dpi} and \ref{eq:dh}), we are able to use the (00)~Einstein equation as a constraint on the model parameters:
\begin{equation}\label{eq:00}
\h^2=\frac{\Omega_m^0}{a^3}+\frac{\Omega_r^0}{a^4} + \frac{\bar  c_2}{6}\h^2\x^2  - 2\bar  c_3\h^4\x^3 
+ \frac{15}{2}\bar c_4\h^6\x^4 - 7 \bar c_5\h^8\x^5 .
\end{equation}
More precisely, we used this constraint both at $z=0$ to fix one of our parameters and, at other redshifts, to check the reliability of our numerical computations (see Sect.~\ref{sec:method}). The parameter we chose to fix at $z=0$ is
\begin{equation}\label{eq:c5}
\bar c_5 = \frac{1}{7}(-1+\Omega_m^0 + \Omega_r^0 + \frac{\bar c_2}{6} - 2 \bar c_3 + \frac{15}{2} \bar c_4) .
\end{equation}
We chose to fix $\bar c_5$ based on the other parameters because allowing it to float introduces significant numerical difficulties when solving equations~\ref{eq:dpi} and \ref{eq:dh}, since it represents the weight of the most non-linear term in these equations. As $\Omega_r^0$ is fixed given $h$, our parameter space has been reduced to $\Omega_m^0, h, \bar c_2,\bar c_3$ and $\bar c_4$.

\subsubsection{Positive energy density}

\par We require that the energy density of the Galileon field be positive from $z=0$ to $z=10^7$ (see \ref{sec:wmap7} and Appendix B). At every redshift 
in this range, this constraint amounts to
\begin{equation}\label{eq:rhopi}
\frac{\rho_{\pi}}{H_0^2 M_P^2}=\frac{\bar c_2}{2}\h^2\x^2  - 6\bar c_3\h^4\x^3 + \frac{45}{2}\bar c_4\h^6\x^4 - 21 \bar c_5\h^8\x^5 > 0 .
\end{equation}
This constraint is not really necessary for generic scalar field models. But as we will see in the following, it has no impact on our analysis because the other theoretical conditions described below are stronger.

\subsubsection{Scalar perturbations}

\par As suggested by \cite{bib:appleby}, outside the quasi-static approximation the propagation equation for $\delta y$ leads to two conditions, which we again checked from $z=0$ to $z=10^7$ to ensure the viability of the linearly perturbed model:
\begin{enumerate}
\item a no-ghost condition, which requires a positive energy for the perturbation
\begin{equation}\label{eq:noghost1}
\kappa_2 + \frac{3}{2} \frac{\kappa_5^2}{\kappa_4}<0 ;
\end{equation}
\item a Laplace stability condition for the propagation speed of the perturbed field
\begin{equation}\label{eq:cs1}
c_s^2=\frac{4\kappa_1\kappa_4\kappa_5 - 2\kappa_3 \kappa_5^2 - 2 \kappa_4^2 \kappa_6}{\kappa_4(2\kappa_4\kappa_2 + 3\kappa_5^2)}>0
\end{equation}
\end{enumerate}
with
\begin{equation}
\kappa_2 = -\frac{\bar c_2}{2}+6\bar c_3\h^2\x - 27\bar c_4 \h^4\x^2 + 30 \bar c_5 \h^6 \x^3.
\end{equation}

\subsubsection{Tensorial perturbations}\label{sec:tensorial}

We also addrd two conditions derived by \citet{bib:felice2011} for the propagation of tensor perturbations. Considering a traceless and divergence-free perturbation $\delta g_{ij} = a^2h_{ij}$, these authors obtained identical perturbed actions at second order for each of the two polarisation modes $h_\oplus$ and $h_\otimes$.  For $h_\oplus$
\begin{equation}
\delta S_T^{(2)}=\frac{1}{2}\int dtd^3xa^3Q_T\left[\dot h^2_\oplus - \frac{c_T^2}{a^2}(\nabla h_\oplus)^2 \right]
\end{equation}
with $Q_T$ and $c_T$ as defined below. From that equation, we extracted two conditions in our parametrization that have to be satisfied (again from $z=0$ to $z=10^7$):
\begin{enumerate}
\item a no-ghost condition:
\begin{equation}\label{eq:noghost2}
\frac{Q_T}{M_P^2} = \frac{1}{2} - \frac{3}{4}\bar c_4 \h^4\x^4 + \frac{3}{2}\bar c_5 \h^5 \x^5 > 0;
\end{equation}
\item a Laplace stability condition:
\begin{equation}\label{eq:cs2}
c_T^2=\frac{\frac{1}{2} +  \frac{1}{4}\bar c_4 \h^4\x^4 + \frac{3}{2}\bar c_5 \h^5 \x^4(\h'\x + \h\x')}{\frac{1}{2} - \frac{3}{4}\bar c_4 \h^4\x^4 + \frac{3}{2}\bar c_5 \h^5 \x^5}>0 .
\end{equation}
\end{enumerate}

These conditions allowed us to reduce our parameter space significantly. The Galileon model contains degeneracies between the $\bar c_i$s, as pointed out in e.g. \cite{bib:barreira}. The above theoretical constraints and our new parametrization allowed us to break degeneracies between the $\bar c_i$ parameters that would make it difficult to converge to a unique best-fit with current cosmological observations. As an example, the tensorial theoretical conditions lead to a significant reduction of  the parameter space (see dark dotted regions in Fig.~\ref{fig:snls}), so that closed probability contours are obtained.

\section{Likelihood analysis method and observables}\label{sec:data}

\par In the following, we define a scenario to be a specific realisation of the cosmological equations for a given set of parameters $\left\lbrace \Omega_m^0,\bar c_2,\bar c_3,\bar c_4 \right\rbrace$.

To perform the likelihood analysis, the method used in \citet{bib:conley} for the analysis of 
SNLS data\footnote{http://casa.colorado.edu/~aaconley/Software.html} was adapted to the Galileon model. For each cosmological probe, a likelihood surface $\mathcal{L}$ was derived by computing the $\chi^2$ for each visited scenario: $\mathcal{L}(\Omega_m^0,\bar c_2, \bar c_3, \bar c_4) \propto e^{-\chi^2/2}$. The way $h$ is treated is described in Sect.~\ref{sec:wmap7}.
Then we report the mean value of the marginalized parameters as the fit values of $\Omega_m^0$ and the $\bar c_i$s.

\subsection{Numerical computation method}\label{sec:method}

To compute numerical solutions to equations \ref{eq:dpi} and \ref{eq:dh}, we used a fourth-order Runge-Kutta method to compute $\h(z)$ and $\x(z)$ iteratively starting from the current epoch, where the initial conditions for $\h$ and $\x$ are specified (see \ref{sec:inicond}), and propagating backwards in time to higher $z$. We used a sufficiently small step size in $z$ to avoid numerical divergences. This is challenging because of the significant non-linearities in our equations. To determine the step size, we therefore required that equation \ref{eq:00}, normalized by $\h^2$, be satisfied at better than $10^{-5}$ for each step. 

At each step of the computation, we also checked that all previously discussed theoretical conditions were satisfied (equations \ref{eq:rhopi}, \ref{eq:noghost1}, \ref{eq:cs1}, \ref{eq:noghost2}, and \ref{eq:cs2}). Cosmological scenarios that fail any of these conditions were rejected and their likelihood set to zero. The result of these requirements is shown e.g. in Fig.~\ref{fig:snls} as dark dotted regions. Equation \ref{eq:rhopi} concerns a negligible number of Galileon scenarios, but the four other constraints lead to a significant reduction of the parameter space. 

\subsection{Data}

Here we describe the cosmological observations we used in our analysis. Special care was taken to choose data that do not depend on additional cosmological assumptions.

\subsubsection{Type Ia supernovae}\label{sec:snls}


The SN~Ia data sample used in this work is the SNLS3 sample described in \cite{bib:conley}. It consists of 472 well-measured supernovae from the SNLS, SDSS, HST, and a variety of low-$z$ surveys.

A type~Ia supernova with intrinsic stretch $s$ and color $\mathcal{C}$ has a rest-frame B-band apparent magnitude $m_B$ that can be modeled as follows:
\begin{equation}
m_B^{mod} = 5 \log_{10} \mathcal{D}_L(z_{hel},z_{CMB},\text{cosmo}) -\alpha(s-1) + \beta.
 \mathcal{C} + \mathcal{M_B},
\end{equation}
where $\mathcal{D}_L$ is the Hubble-constant free luminosity distance, which in a flat Universe is given by
\begin{equation}
\mathcal{D}_L(z_{hel},z_{CMB},\text{cosmo}) = (1+z_{hel})\int_0^{z_{CMB}} \frac{dz}{\bar H(z,\mathrm{cosmo})}.
\end{equation}
$z_{hel}$ and $z_{CMB}$ are the SN~Ia redshift in the heliocentric and CMB rest frames, respectively, "cosmo" represents the cosmological parameters of the model. $\alpha$ and $\beta$ are parameters describing the light-curve width-luminosity and
color-luminosity relationships for SNe~Ia. $\mathcal{M_B}$ is defined as $\mathcal{M_B}=M_B+5\log_{10} c/H_0+25$, where $M_B$ is the rest-frame absolute magnitude of a fiducial ($s=1, \mathcal{C} = 0$) SN Ia in the B-band, and $c/H_0$ is expressed in Mpc. $\alpha,\beta$ and $\mathcal{M_B}$ are nuisance parameters that are fit simultaneously with the cosmological parameters. As in \cite{bib:conley} and \cite{bib:sullivan}, we allowed for different $\mathcal{M_B}$ in galaxies with the host galaxy stellar mass below and above $10^{10}$ M$_\odot$ to account for relations between SN Ia brightness and host properties that are not corrected for via the standard $s$ and $\mathcal{C}$ relations. When computing type Ia supernova distance luminosities in Sect.~\ref{sec:results}, we neglect the radiation component in $\bar H(z)$, since all measurements are restricted to redshifts below 1.4 where the effects of radiation density are negligible.

Systematic uncertainties must be treated carefully when using SN Ia data, because they depend on $\alpha$ and $\beta$ and due to covariances between different supernovae. We followed the treatment of \cite{bib:conley} and \cite{bib:sullivan}.


\subsubsection{Cosmological microwave background}\label{sec:wmap7}

\par The CMB is a powerful probe to constrain the expansion history of the Universe because it gives high-redshift cosmological observables. The power spectrum provides much information on the content of the Universe and the relations between the different fluids, as long as we are able to model the thermodynamics of these fluids before recombination. The Galileon model does not modify the standard baryon-photon flux physics as long as the Galileon field does not couple directly to matter, as is assumed in this work. Thus, the usual formulae and predictions used in the standard analysis of the CMB power spectrum remain valid.

The positions of the acoustic peaks can be quantified by three observables: $\left\lbrace l_a, R, z_* \right\rbrace$ (see e.g.  \cite{bib:komatsu11} and \cite{bib:komatsu09}), where $l_a$ is the acoustic scale related to the comoving sound speed horizon, $R$ is the shift parameter related to the distance between us and the last scattering surface, and $z_*$ is the redshift of the last scattering surface. These quantities are derived from the angular diameter distance, which in a flat space is given by
\begin{equation}
D_A(z)=\frac{c}{H_0}\frac{1}{1+z}\int_0^z\frac{dz'}{\bar H(z')},
\end{equation}
and from the comoving sound speed horizon:
\begin{equation}\label{eq:rs}
r_s(z)=\frac{c}{H_0}\int_0^\frac{1}{1+z}da\frac{\bar c_s(a)}{a^2\bar H(a)} .
\end{equation}
$\bar c_s$ is the usual normalized sound speed in the baryon-photon fluid before recombination:
\begin{equation}
\bar c_s = \frac{1}{\sqrt{1+3(3\Omega_b^0/4\Omega_\gamma^0)a}},
\end{equation}
where $\Omega_b^0$ is the baryon energy density parameter today.

With the above definitions, the acoustic scale $l_a$ is given by
\begin{equation}\label{eq:la}
l_a = (1+z_*)\frac{\pi D_A(z_*)}{r_s(z_*)},
\end{equation}
and the shift parameter $R$ by
\begin{equation}
R = \frac{\sqrt{\Omega_m^0H_0^2}}{c}(1+z_*)D_A(z_*)=\sqrt{\Omega_m^0}\int_0^z\frac{dz'}{\bar H(z')}.
\end{equation}
$z_*$ is given by the fitting formula of \cite{bib:hu}:
\begin{equation}\label{eq:zstar}
z_*=1048\left[1+0.00124(\Omega_b^0h^2)^{-0.738}\right]
\left[1+g_1(\Omega_m^0h^2)^{g_2}\right]
\end{equation}
\begin{align}
g_1=\frac{0.0783(\Omega_b^0h^2)^{-0.238}}{1+39.5(\Omega_b^0h^2)^{0.763}}\\
g_2=\frac{0.560}{1+21.1(\Omega_b^0h^2)^{1.81}}.
\end{align}
According to \cite{bib:hu}, formula \ref{eq:zstar} is valid for a wide range of $\Omega_m^0h^2$ and $\Omega_b^0h^2$.

To compare these observables with the seven-year WMAP data (WMAP7), we followed the numerical recipe given in \cite{bib:komatsu09}. The key point of this recipe is that for each cosmological scenario, $\chi^2_{CMB}$ must be minimized over $h$ and $\Omega_b^0h^2$, which appear in equation \ref{eq:zstar} and in the computation of $\bar H(z)$ through $\Omega_r^0$ (see equation \ref{eq:lambda}).

An important feature to note is that we have to solve equations \ref{eq:dpi} and \ref{eq:dh} from $a=1$ to $a=0$ to compute the CMB observables. Numerically, however, we cannot reach $a=0$ ($z=\infty$) because of numerical divergences.  To avoid them, we carried out these computations up to $a=10^{-7}$ and then linearly extrapolated the value of the integral to $a=0$ (for more details on the reliability of this approximation see Appendix B). Thus, the theoretical constraints of \ref{sec:theoconstraints} were checked from $a=1$ to $a=10^{-7}$.

Finally, 
because CMB observables depend explicitly on $H_0$, we imposed a Gaussian prior on its value, $h = 0.737 \pm 0.024$ as measured by \cite{bib:riess11} from low-redshift SNe~Ia and Cepheid variables.

The WMAP7 recommended best-fit values of the CMB observables are
\begin{equation}
\langle \mathbf{V}_{CMB}\rangle=\left( \begin{array}{c}  \langle l_a \rangle \\ \langle R \rangle \\ \langle z_* \rangle \end{array} \right)
= \left( \begin{array}{c} 302.09 \pm 0.76 \\ 1.725 \pm 0.018 \\ 1091.3 \pm 0.91 \end{array} \right),
\end{equation}
with the corresponding inverse covariance matrix:
\begin{equation}
 \mathbf{C}^{-1}_{CMB}=\left( \begin{array}{ccc} 2.305 & 29.698 & -1.333 \\ 29.698 & 6825.270 & -113.180 \\ -1.333 & -113.180 & 3.414 \end{array} \right)
\end{equation}
from \citet{bib:komatsu11}. As pointed out by \citet{bib:nesseris}, the uncoupled Galileon model fulfils the assumptions required in \citet{bib:komatsu09} to use these distance priors, namely a FLRW Universe with the standard number of neutrinos and a dark energy background with negligible interactions with the primordial Universe. Once the observables $\left\lbrace l_a,R,z_* \right\rbrace$ were computed in a cosmological scenario, we built the difference vector:
\begin{equation}
\Delta \mathbf{V}_{CMB}= \left( \begin{array}{c} l_a \\  R \\  z_* \end{array} \right) -  \langle\mathbf{V}_{CMB}\rangle
\end{equation}
and computed the CMB contribution to the total $\chi^2$ :
\begin{equation}
\chi^2_{CMB+H_0} = \Delta  \mathbf{V}_{CMB}^T  \mathbf{C}^{-1}_{CMB} \Delta  \mathbf{V}_{CMB} +\frac{(h-0.738)^2}{0.024^2}.
\end{equation}

\subsubsection{Baryonic acoustic oscillations}\label{sec:bao}

\par BAO distances provide information on the imprint of the comoving sound horizon after recombination on the distribution of galaxies. The BAO observable is defined as $y_s(z)=r_s(z_d)/D_V(z)$, where $r_s$ is the comoving sound horizon at the baryon drag epoch redshift $z_d$, and $D_V(z)$ is the effective distance \citep{bib:eisenstein05} given by
\begin{equation}
D_V(z)=\left[(1+z)^2D_A^2(z)\frac{cz}{H(z)}\right]^{1/3} .
\end{equation}
$z_d$ is computed using the \cite{bib:eisenstein98} fitting formula:
\begin{equation}
z_d=\frac{1291(\Omega_m^0h^2)^{0.251}}{1+0.659(\Omega_m^0h^2)^{0.828}}
\left[1+b_1(\Omega_b^0h^2)^{b_2}\right]
\end{equation}
\begin{align}
b_1 &= 0.313(\Omega_m^0h^2)^{-0.419}
\left[1+0.607(\Omega_m^0h^2)^{0.674}\right] \\
b_2 &= 0.238(\Omega_m^0h^2)^{0.223} .
\end{align}
This formula remains valid for a Galileon field not coupled to matter.

Therefore BAO distances depend on $h$ and $\Omega_b^0$ as the CMB observables so we followed the same recipe as previously mentioned to compute them, including the $H_0$ prior from \cite{bib:riess11}. We also made the same approximation as for the CMB to compute $r_s$. The minimization over $h$ and $\Omega_b^0h^2$ was performed independently for CMB and BAO when their individual constraints are derived and simultaneously when combined constraints were computed.

We used the dataset of distances derived from galaxy surveys as published in the SDSS-III BOSS cosmological analysis (\cite{bib:boss12} and \cite{bib:sanchez}) to avoid redshift overlaps in the measurements (see Table~\ref{tab:bao}). 


\begin{table*}[htb]
\caption[]{BAO measurements.}
\label{tab:bao}
\begin{center}
\begin{tabular}{cccc} \hline \hline \\ [-1.ex]
$z$  & $y_s^{mes}(z)$ & Survey & Reference \\ [1ex] \hline  \\ [-1.ex]
0.106 & $0.336 \pm 0.015$ & 6dFGS & \cite{bib:beutler} \\
0.35 & $0.1126 \pm 0.0022$ & SDSS LRG & \cite{bib:padmanabhan} \\
0.57 & $0.0732 \pm 0.0012$ & BOSS CMASS & \cite{bib:boss12} \\  [1ex] \hline 
\end{tabular}
\end{center}
\end{table*}

For a cosmological constraint derived from BAO distances alone, the BAO contribution to the total $\chi^2$ is given by
\begin{equation}\label{eq:chi2bao}
\begin{split}
\chi^2_{BAO+H_0} = \sum_{z} \frac{(y_s(z)-y_s^{mes}(z))^2}{\sigma_{y_s}^2}+\frac{(h-0.738)^2}{0.024^2} \\ +\frac{(\Omega_b^0h^2-0.02249)^2}{0.00057^2},
\end{split}
\end{equation}
where we added a Gaussian prior on $\Omega_b^0h^2$ when dealing with this probe alone.

When BAO and CMB probes were combined, we computed their contributions to the $\chi^2$ simultaneously to avoid over-counting the Hubble constant
prior. Therefore, the combined contribution is
\begin{equation}
\begin{split}
\chi^2_{CMB+BAO+H_0} = \Delta \mathbf{V}_{CMB}^T  \mathbf{C}^{-1}_{CMB} \Delta  \mathbf{V}_{CMB}\\ + \sum_{z} \frac{(y_s(z)-y_s^{mes}(z))^2}{\sigma_{y_s}^2} +\frac{(h-0.738)^2}{0.024^2} .
\end{split}
\end{equation}

\subsubsection{Growth rate of structures}\label{sec:gof}

\begin{table*}[htb]
\caption[]{Growth data.}
\label{tab:gof}
\begin{center}
\begin{tabular}{cccccc} \hline \hline \\ [-1ex]
$z$ & $f\sigma_8(z)$ & $F(z)$ & $r$ & Survey & Reference \\  [1ex] \hline \\ [-1ex]
0.067 & $0.423\pm 0.055$ & - & - & 6dFGRS (a) & \cite{bib:beutler12} \\ [1ex]
0.17 & $0.51\pm 0.06$ & - & - & 2dFGRS (a) & \cite{bib:percival04} \\ [1ex]
0.22 & $0.53\pm 0.14$ & $0.28\pm0.04$ & 0.83 & WiggleZ & \cite{bib:blake11b} \\ [1ex] 
0.25 & $0.351\pm 0.058$ & - & - & SDSS LRG (b) & \cite{bib:samushia12a} \\ [1ex] 
0.37 & $0.460\pm 0.038$ & - & - & SDSS LRG (b) & \cite{bib:samushia12a} \\ [1ex] 
0.41 & $0.40\pm 0.13$ & $0.44\pm0.07$ & 0.94 & WiggleZ & \cite{bib:blake11b} \\ [1ex] 
0.57 & $0.430\pm 0.067$ & $0.677\pm0.042$ & 0.871 & BOSS CMASS & \cite{bib:reid} \\ [1ex] 
0.6 & $0.37\pm 0.08$ & $0.68\pm0.06$ & 0.89 & WiggleZ & \cite{bib:blake11b} \\ [1ex] 
0.78 & $0.49\pm 0.12$ & $0.49\pm0.12$ & 0.84 & WiggleZ & \cite{bib:blake11b} \\ [1ex]  \hline \\ [-1ex]
\end{tabular}
\tablefoot{$r$ is the cross-correlation in $(F,f\sigma_8)$. (a) Alcock-Paczynski effect is negligible at low redshift. 
(b) Values of $f\sigma_8$ are corrected for the Alcock-Paczynski effect but no $F(z)$ values are provided.}
\end{center}
\end{table*}

\par The cosmological growth of structures is a critical test of the Galileon model, as noted by many authors (see \cite{bib:linder05} for example). It is a very discriminant constraint for distinguishing dark energy and modified gravity models. Many models can mimic $\Lambda$CDM behavior for the expansion history of the Universe, but all modify gravity and structure formation in a different manner. 

In linear perturbation theory, the growth of a matter perturbation $\delta_m = \delta \rho_m/ \rho_m$ is governed by the equation
\begin{equation}
\ddot{\delta_m} + 2H\dot{\delta_m}-4\pi G_N \rho_m \delta_m=0 .
\end{equation}
But as argued in \cite{bib:linder05} and as used in \cite{bib:komatsu09}, it is better to study the growth evolution with the function $g(a)\equiv D(a) / a \equiv \delta_m(a)/(a\delta_m(1))$. In the Galileon case, the Newton constant is replaced by $G_{\mathrm{eff}}^{(\psi)}(a)$ as given in equation \ref{eq:geff}. The $g(a)$ is obtained by solving the following second-order differential equation
\begin{multline}\label{eq:growth}
\frac{d^2g}{da^2} + \frac{1}{a}\left(5+\frac{a}{\bar H}\frac{d\bar H}{da}\right)\frac{dg}{da} \\
 +\frac{1}{a^2}\left(3+\frac{a}{\bar H}\frac{d\bar H}{da}-\frac{3}{2}\frac{G_{\mathrm{eff}}^{(\psi)}}{G_N}\frac{\Omega_m^0}{a^3 \bar H^2}\right) = 0 .
\end{multline}
A natural choice for the initial conditions is $g(a_{\mathrm{initial}})=1$ and $dg/da\mid_{a_{\mathrm{initial}}}=0$ \citep{bib:komatsu09}, where $a_{\mathrm{initial}}$ is $0.001\approx 1/(1+z_{*})$. We checked that our results do not depend on this choice as long as $a_{\mathrm{initial}}$ is taken between  $10^{-2}$ and $10^{-5}$.

Measurements of the rate of growth of cosmic structures from redshift space distortions can be expressed in terms of $f(a)=d \ln D(a) / d \ln a$ or $f\sigma_8(a)$, where $\sigma_8$ is the normalization of the matter power spectrum. $f\sigma_8(a)$ is known to be less sensitive to the overall normalization of the power spectrum model used to derive the measurements \citep{bib:song}. Accordingly this is the observable we chose in this work. To predict $f\sigma_8(a)$ in our analysis, we solved equation~\ref{eq:growth} to obtain $g(a)$, from which we deduced $f(a)$ and $D(a)$, and we computed $\sigma_8(a)$ in the following way \citep{bib:samushia12a}:
\begin{equation}\label{eq:s81}
\sigma_8(a) = \sigma_8(a_{\mathrm{initial}})\frac{D(a)}{D(a_{\mathrm{initial}})},
\end{equation}
where
\begin{equation}\label{eq:s82}
\sigma_8(a_{\mathrm{initial}}) = \sigma_8^{\mathrm{WMAP7}}(a=1)\frac{D^{\Lambda\mathrm{CDM}}(a_{*})}{D^{\Lambda\mathrm{CDM}}(a=1)},
\end{equation}
and $\sigma_8^{\mathrm{WMAP7}}(a=1)=0.811^{+0.030}_{-0.031}$ is the present value of the CMB power spectrum normalization published by \cite{bib:komatsu11} in the framework of the $\Lambda$CDM model. Equation~\ref{eq:s82} states that the normalization of the CMB power spectrum at decoupling is the same in the $\Lambda$CDM and Galileon models, which is consistent with our assumption that the CMB physics is not modified by the Galileon presence. This equation holds  if $D(a)$ has no scale dependence, which is the case  in both models in the linear regime. Equation~\ref{eq:s81} takes into account the different growth histories since recombination in the two models. 

However, stand-alone $f\sigma_8(a)$ measurements extracted from observed matter power spectra usually use a fiducial cosmology, which assumes General Relativity. This hypothesis is no longer necessary when taking into account the Alcock-Paczynski effect \citep{bib:alcock} in the power spectrum analysis. This results in joint measurements of $f\sigma_8(a)$ and the Alcock-Paczynski parameter $F(a)\equiv c^{-1}D_A(a)H(a)/a$, which are to be preferred when constraining modified gravity models (see e.g. \cite{bib:beutler12} and \cite{bib:samushia12b}). Note that equations \ref{eq:dpi} and \ref{eq:dh} are all we need to predict $F(a)$ in the Galileon model.

The measurements of $f\sigma_8(z)$ and $F(z)$ used in this work are summarized in Table~\ref{tab:gof}. To compare these with our model, we first solved equations \ref{eq:dpi} and \ref{eq:dh} from $a=1$ to $a_{\mathrm{initial}}$ to obtain values of $\bar H(a)$, $F(a)$ and $G_{\mathrm{eff}}^{(\psi)}(a)/G_N$, and then solved equation \ref{eq:growth} from $a_{\mathrm{initial}}$ to $a=1$, which provides us with $f\sigma_8(z)$ predictions. 

Because $F(z)$ and $f\sigma_8(z)$ measurements are correlated, a covariance matrix $\mathbf{C}_{\mathrm{GoS}}$ was built using data presented in Table~\ref{tab:gof}. Moreover, our $f\sigma_8$ prediction relies on the WMAP7 measurement of $\sigma_8(a=1)$ (equation \ref{eq:s82}), so the WMAP7 experimental uncertainty is also propagated to the diagonal and off-diagonal terms of $\mathbf{C}_{\mathrm{GoS}}$. Then a vector $\mathbf{V}_{GoS}$ containing all predictions at each $z_i$ was built
\begin{equation}
\mathbf{V}_{GoS}=\left( \begin{array}{c}  \vdots \\ f\sigma_8(z_i) \\ F(z_i) \\ \vdots \end{array} \right).
\end{equation}
The contribution of the growth rate of structures to the total $\chi^2$ is then
\begin{equation}
\chi^2_{GoS}=\Delta  \mathbf{V}_{GoS}^T  \mathbf{C}^{-1}_{GoS} \Delta  \mathbf{V}_{GoS},
\end{equation}
with $\Delta  \mathbf{V}_{GoS}=\mathbf{V}_{GoS}-\langle \mathbf{V}_{GoS}\rangle$, where $\langle \mathbf{V}_{GoS}\rangle$ contains the measurements of Table~\ref{tab:gof}.

Note that equation \ref{eq:lambda} requires a value for $\Omega_r^0$, and hence in principle this equation should be simultaneously solved with the BAO and CMB constraints using the same prior on $H_0$. However, we found that this has essentially no effect on our $\chi^2$. Therefore, we set here $h$ to the value derived from the $H_0$ measurements of \cite{bib:riess11} to accelerate the computation.

\section{Results}\label{sec:results}

\par In the following we present the results of the experimental constraints on the Galileon model derived from the cosmological probes.

\subsection{SN constraints}\label{sec:snlsconstraints}

\par Results from SN~Ia data are presented in Fig.~\ref{fig:snls} and Table~\ref{tab:snls}.

\subsubsection{SN results}

Despite the large number of free parameters in the model, we obtained closed probability contours in any two-dimensional
projection of the parameter space.  We observed strong correlations between the $\bar c_i$s, especially between $\bar c_2$ and $\bar c_3$.

We note that the best-fit value for $\Omega_m^0\approx 0.27$ is compatible with the current constraints obtained in the $\Lambda$CDM or FWCDM models. The $\bar c_i$s are found to be globally of the order of $\approx -1$. From the best-fit values of the parameters, we derived the value of $\bar c_5$ using equation \ref{eq:c5} and find 
$\bar c_5=-0.349^{+0.632}_{-0.555}$, including systematic uncertainties.

In the following we discuss the impact of fixing the nuisance parameters $\alpha$ and $\beta$ and the effect of systematics on the best-fit values.

\subsubsection{Impact of nuisance parameters}\label{sec:nuisance}

\begin{figure}[hbtp]
\begin{center}
\epsfig{figure=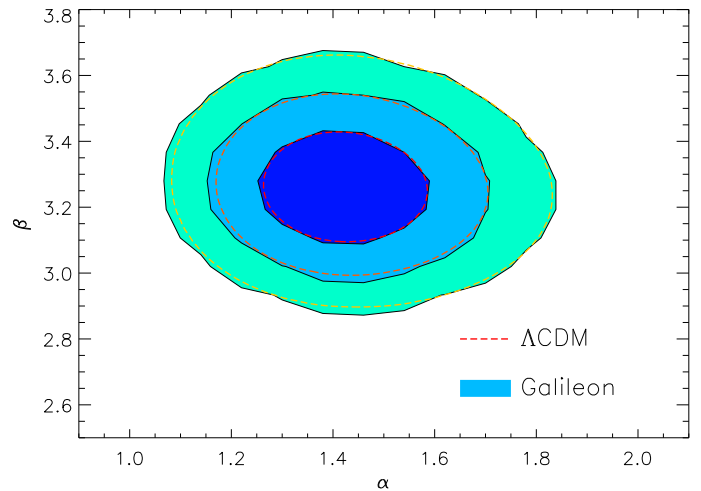, width=\columnwidth} 
\caption[]{Confidence contours for the SN nuisance parameters $\alpha$ and $\beta$ when marginalizing over all other parameters of the model. Dashed red contours represent 68.3\%, 95.4\%, and 99.7\% probability contours for the $\Lambda$CDM model. Filled blue contours are for the Galileon model. Note that they are nearly identical, the Galileon one is just 2.8\% wider, which is likely due to 
larger steps in $\alpha$ and $\beta$. See Table~\ref{tab:snls} for numerical values.} 
\label{fig:alphabeta}
\end{center}
\end{figure}

\begin{figure*}[hbtp]
\begin{center}
\epsfig{figure=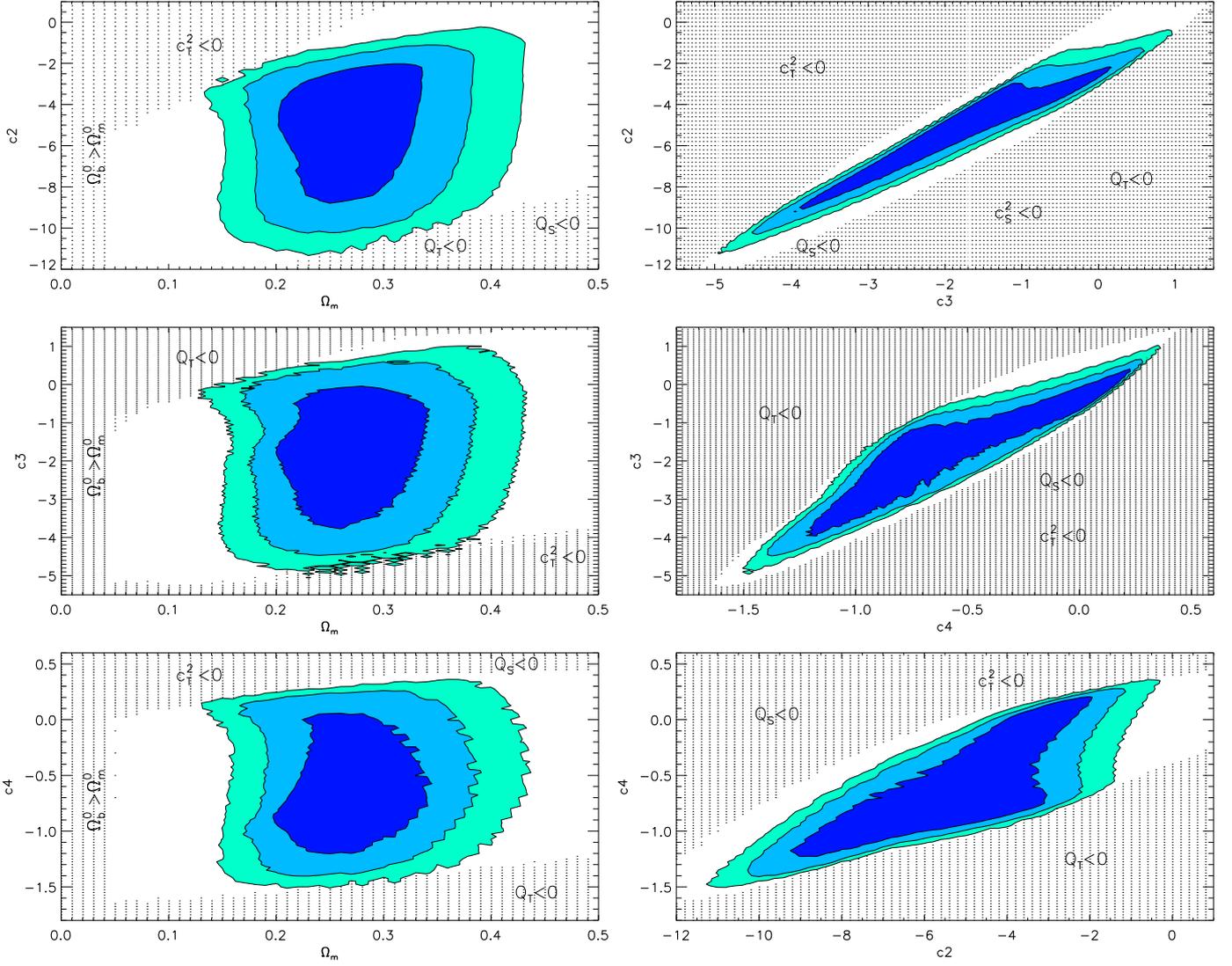, width=\textwidth} 
\caption[]{Experimental constraints on the Galileon model from SNLS3 data alone. 
To represent the four-dimensional likelihood $\mathcal{L}(\Omega_m^0,\bar c_2, \bar c_3, \bar c_4)$, six two-dimensional contours for each pair of the Galileon model parameters are presented, after marginalizing over $\mathcal{M}_B^1$, $\mathcal{M}_B^2$, $\alpha$, $\beta$, and the remaining Galileon parameters. The filled dark, medium, and light-blue contours enclose 68.3, 95.4, and 99.7\% of the probability, respectively. The contours include statistical and all identified systematic uncertainties. The dark dotted regions correspond to scenarios rejected by theoretical constraints, as described in the text. Labels in these regions indicate the main cause for excluding the scenarios.} 
\label{fig:snls}
\end{center}
\end{figure*}

\begin{table*}[htbp]
\caption[]{Cosmological constraints on the Galileon model from the SNLS3 sample}
\label{tab:snls}
\begin{center}
\begin{tabular}{cccccccccc} \hline \hline \\ [-1ex]
Method & $\Omega_m^0$ & $\bar c_2$ & $\bar c_3$ & $\bar c_4$ & $\alpha$ & $\beta$ & $\mathcal{M}_B^1$ & $\mathcal{M}_B^2$ & $\chi^2$ \\  [1ex] \hline \\ [-1ex]
Stat+sys+$\alpha\beta$ & $0.273^{+0.057}_{-0.042}$ & $-5.235^{+1.875}_{-2.767}$ & $-1.779^{+1.073}_{-1.416}$ & $-0.587^{+0.515}_{-0.349}$ &  $1.428^{+0.121}_{-0.098}$ & $3.263^{+0.121}_{-0.103}$ & 23.997 & 23.950 & 415.4 \\ [1ex] \hline \\ [-1ex] 
Stat+sys & $0.273^{+0.054}_{-0.042}$ & $-5.240^{+1.880}_{-2.802}$ & $-1.781^{+1.071}_{-1.426}$ & $-0.588^{+0.516}_{-0.348}$ & 1.428 & 3.263 & 23.997 & 23.950 & 420.1\\ [1ex]  \hline \\ [-1ex] 
Stat only & $0.294^{+0.045}_{-0.039}$ & $-4.765^{+1.725}_{-2.921}$ & $-1.586^{+0.987}_{-1.474}$ & $-0.541^{+0.502}_{-0.338}$ & 1.451 & 3.165 & 24.022 & 23.951 & 441.8\\ [1ex]  \hline  
\end{tabular}
\tablefoot{Results were computed using either statistical and systematic uncertainties combined, or statistical uncertainties only. In the first line, we marginalized over $\alpha$ and $\beta$, whereas in the last two lines, $\alpha$ and $\beta$ were kept fixed to their marginalized values. No errors are given on $\mathcal{M}_B^1$ and $\mathcal{M}_B^2$ because they were analytically marginalized over (see \cite{bib:conley}).}
\end{center}
\end{table*}

\begin{figure*}[hbtp]
\begin{center}
\epsfig{figure=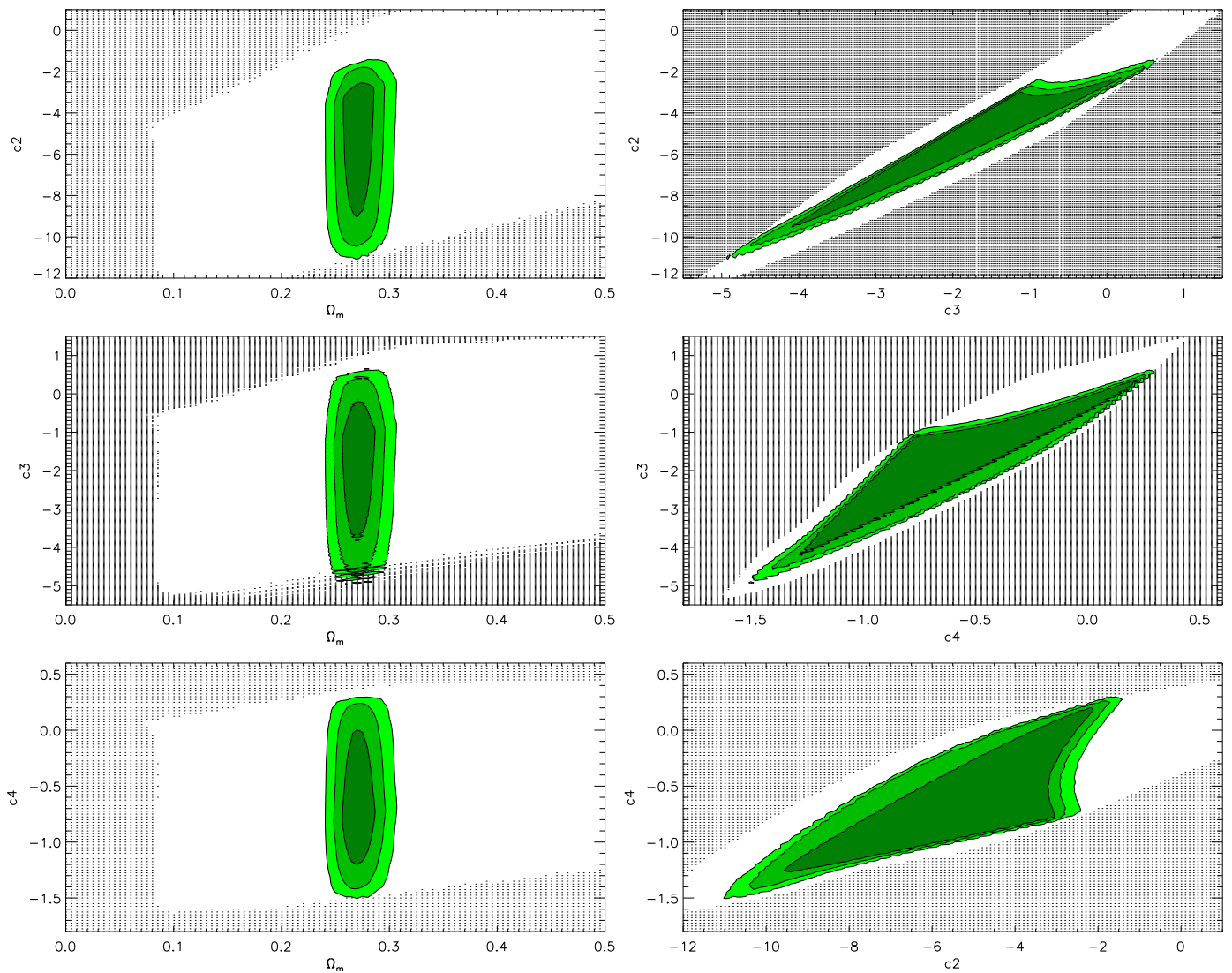, width=\textwidth} 
\caption[]{Experimental constraints on the Galileon model from WMAP7+BAO+H0 data. To represent the four-dimensional likelihood $\mathcal{L}(\Omega_m^0,\bar c_2, \bar c_3, \bar c_4)$, six two-dimensional contours for each pair of the Galileon model parameters are presented after marginalizing over the left over Galileon parameters. The filled dark, medium, and light-green contours enclose 68.3, 95.4, and 99.7\% of the probability, respectively. Dark dotted regions correspond to scenarios rejected by theoretical constraints.}
\label{fig:baowmap7}
\end{center}
\end{figure*}

When marginalizing over the cosmological parameters, the best-fit values of the SN nuisance parameters $\alpha$, $\beta$, $\mathcal{M}_B^1$, and $\mathcal{M}_B^2$ in the Galileon context are identical to those published for the $\Lambda$CDM model, as shown in Fig.~\ref{fig:alphabeta} and Table~\ref{tab:snls}. This is a truly important point to note. It means that the modeling of the SN~Ia physics contained in these nuisance parameters is adequate for these two cosmological models despite their differences.

In principle, the correct method to use when analyzing SN~Ia data is to scan and marginalize over the nuisance parameters. However, once the best-fit values of $\alpha$ and $\beta$ are known, keeping them fixed to their best-fit values in any study using the same SN sample has a negligible impact on our results (see Table~\ref{tab:snls}). In the Galileon case, the contour areas decrease by only 0.7\% and have the same shape as in Fig.~\ref{fig:snls}. For future studies with the SNLS3 sample in the $\Lambda$CDM or Galileon models, our analysis therefore demonstrates that it is reasonable to keep the nuisance parameters fixed to the values published the SNLS papers.

\subsubsection{Impact of systematic uncertainties}\label{sec:statsys}



From the results in Table~\ref{tab:snls}, we note that the identified systematic uncertainties shift the best-fit values of the Galileon parameters by less than their statistical uncertainties.
With systematics included, the area of the inner contours increases by about 53\%. This is less than what is observed in fits to the 
$\Lambda$CDM or FWCDM models (103\% and 80\% respectively, see \cite{bib:conley}).


\subsection{Combined CMB, BAO, and $H_0$ constraints}

The results using CMB, BAO, and $H_0$ data are presented in Fig.~\ref{fig:baowmap7} and Table~\ref{tab:results}.

The combined WMAP7+BAO+H0 data provide a very powerful constraint on $\Omega_m^0$, but no tighter constraints on the $\bar c_i$ than SNe~Ia alone. $\Omega_m^0=0.272^{+0.014}_{-0.009}$ is, as for the SNLS3 sample, close to the current best estimates for this parameter in the standard cosmologies, but this time with very sharp error bars competitive with the most recent studies on other cosmological models. However, the $\bar c_i$ best-fit values are similar to those predicted with the SNLS3 sample.

\begin{figure}[hbtp]
\begin{center}
\epsfig{figure=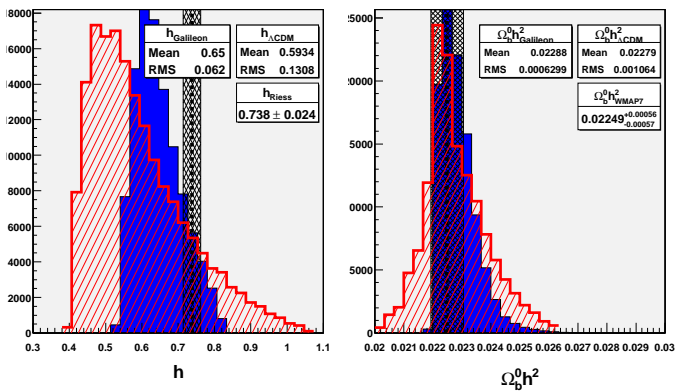, width=\columnwidth} 
\caption[]{Minimized values of $h$ and $\Omega_b^0h^2$ for a large subset of tested scenarios, in $\Lambda$CDM (red dashed histogram) and in the Galileon cosmology (blue filled histogram). Dashed black bands represent the measurements of $H_0$ from \cite{bib:riess11} and $\Omega_b^0h^2$ from \cite{bib:komatsu11}. Only scenarios with $\chi^2<200$ enter these histograms to deal only with pertinent scenarios. Note that both models give values of $h$ and $\Omega_b^0h^2$ that agree with the measurements.
}\label{fig:h0obh2}
\end{center}
\end{figure}

To use the WMAP7+BAO+H0 data, $h$ and $\Omega_b^0h^2$ have to be minimized for each explored Galileon scenario. Minimized values of these parameters are collected in the histograms of Fig.~\ref{fig:h0obh2} for the subset of the scenarios that fulfilled the theoretical constraints. Values for the best-fit scenarios are reported in Table~\ref{tab:results}. For the Galileon model, the $h$ distribution has a mean of 0.65 with a dispersion of 0.06, compatible with the $H_0$ prior. The constraint on $h$ is slightly lower than the \cite{bib:riess11} value, but the same behavior is obtained for the $\Lambda$CDM model using the same program and data. The central value for the $\Omega_b^0h^2$ distribution is fully compatible with the WMAP7 value, for the Galileon and the $\Lambda$CDM model. However, in the Galileon model, values below 0.22 are much more disfavored.

\begin{figure}[hbtp]
	\epsfig{figure=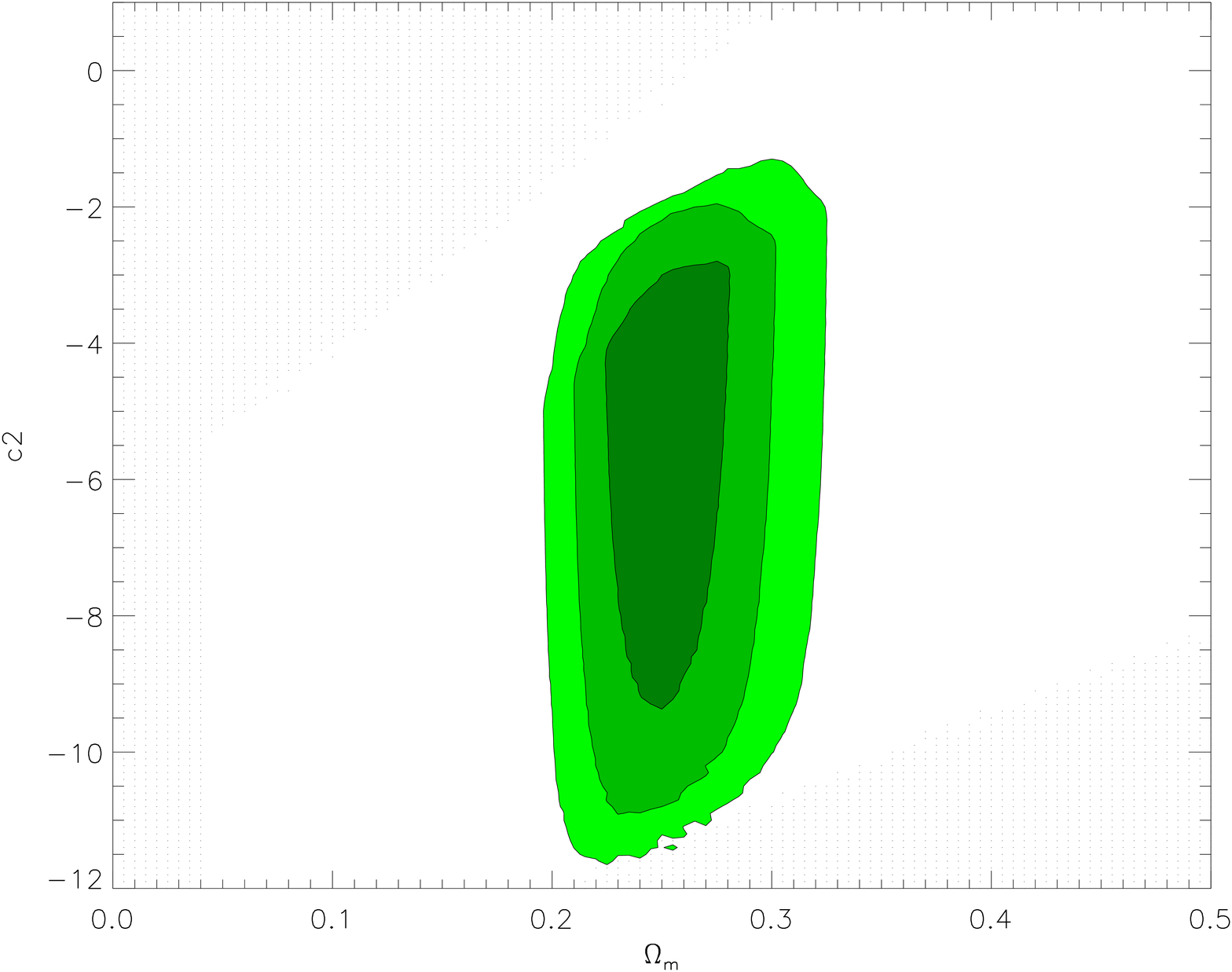, width=1\columnwidth} 
	\epsfig{figure=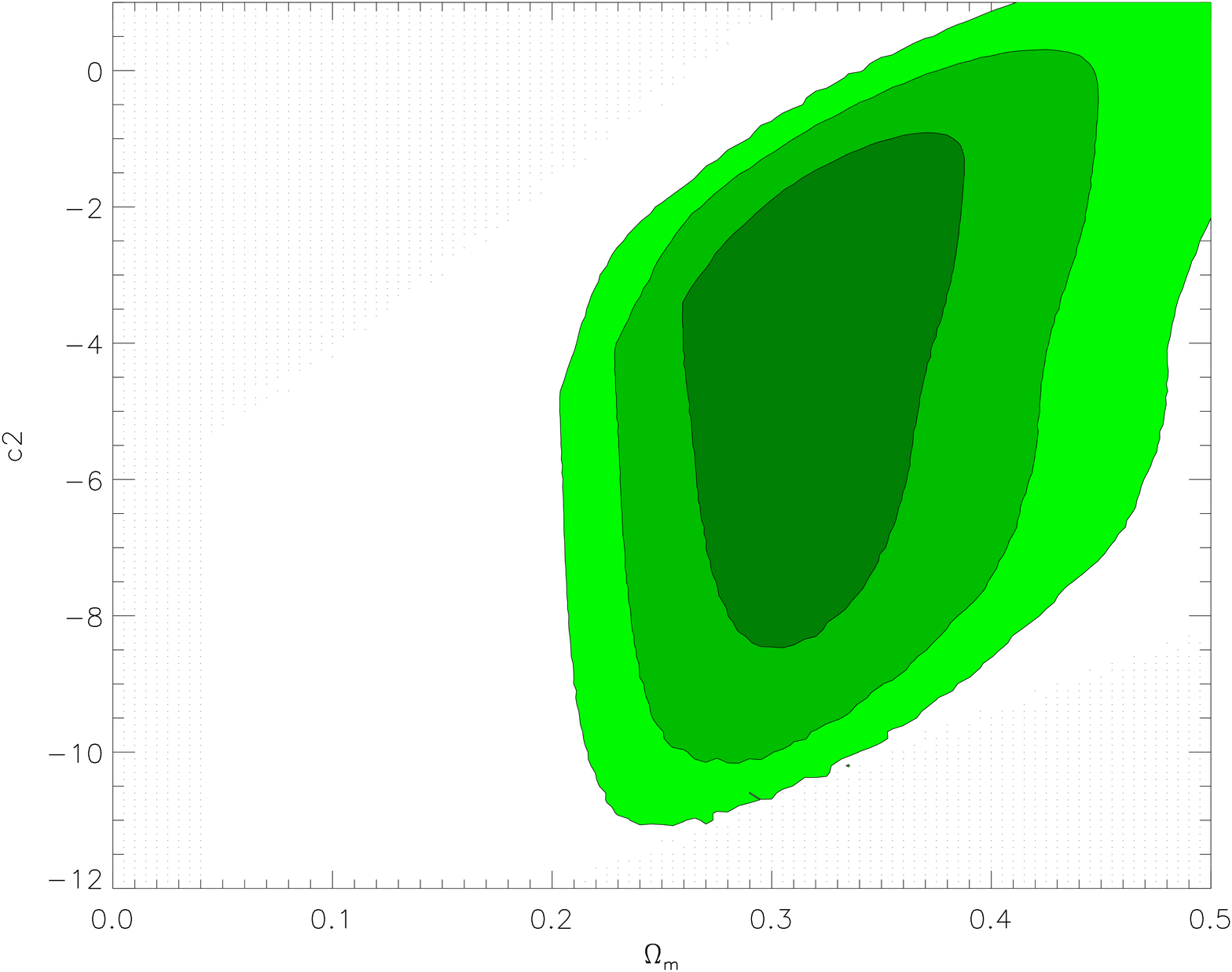, width=1\columnwidth} 
\caption[]{Experimental constraints on the Galileon parameters $\Omega_m^0$ and $\bar c_2$ from WMAP7+H0 data (top panel) and from BAO+H0 data (bottom panel). 
The four-dimensional likelihood $\mathcal{L}(\Omega_m^0,\bar c_2, \bar c_3, \bar c_4)$ has been marginalized over $\bar c_3$ and $\bar c_4$. The filled dark, medium, and light-green contours enclose 68.3, 95.4, and 99.7\% of the probability, respectively. Dark dotted regions correspond to scenarios rejected by theoretical constraints. A Gaussian prior on $\Omega_b^0h^2$ based on the WMAP7 value has been added to the BAO+H0 fit.}\label{fig:bao+wmap7}
\end{figure}

For completeness, we present in Fig.~\ref{fig:bao+wmap7} examples of results obtained from the WMAP7+H0 and BAO+H0 probes separately. Both plots were obtained with a minimization on $h$ and $\Omega_b^0h^2$, but a Gaussian prior on $\Omega_b^0h^2$ was added for the BAO (see Equation~\ref{eq:chi2bao}). We used the WMAP7 constraint for that prior because Fig.~\ref{fig:h0obh2} shows that the Galileon model is consistent with it.


\subsection{Growth-of-structure constraints}\label{sec:growth}

\begin{figure*}[hbtp]
\begin{center}
\epsfig{figure=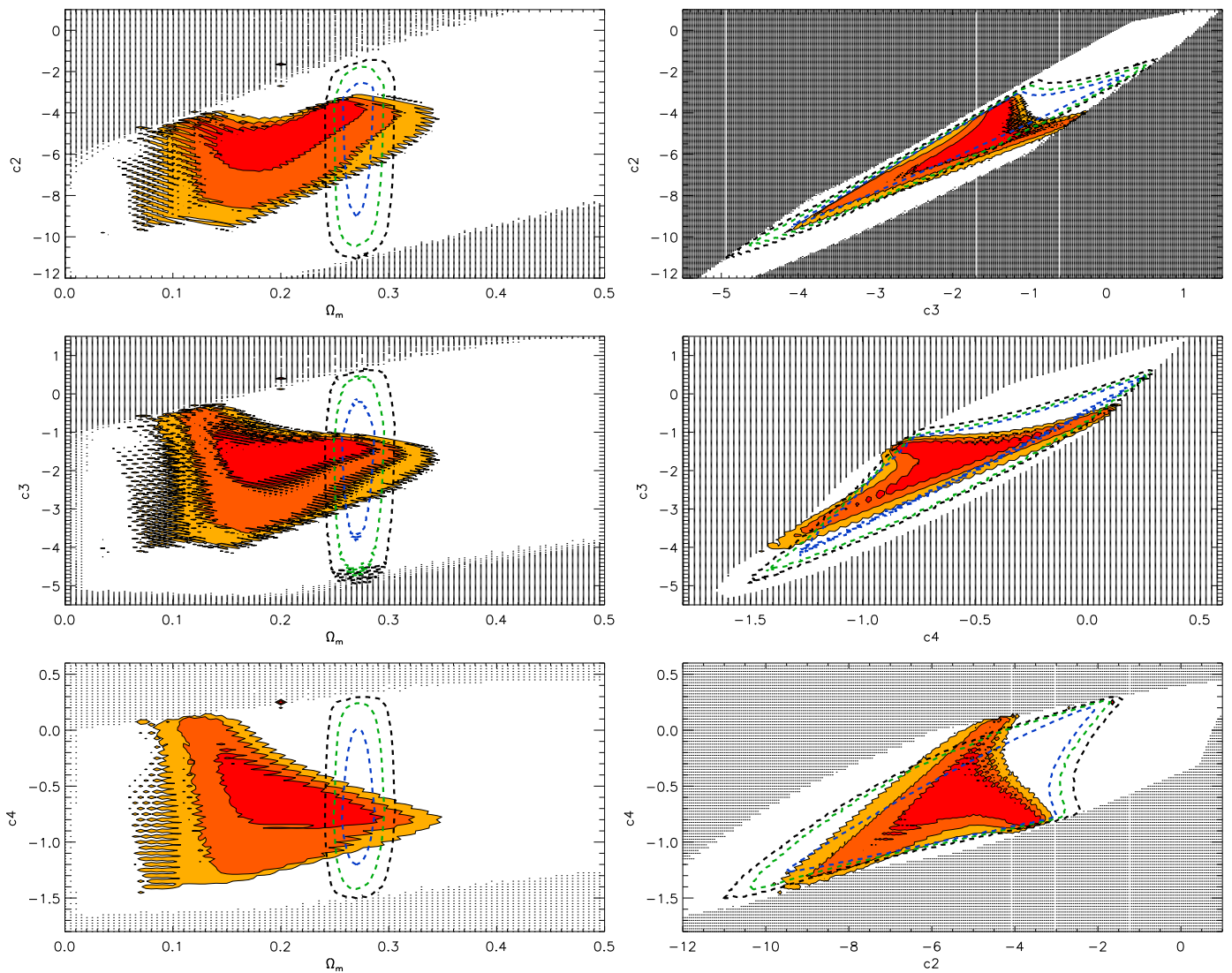, width=\textwidth} 
\caption[]{Experimental constraints on the Galileon model from growth data (red) and from SNLS3+WMAP7+BAO+H0 combined constraints (dashed). The filled dark, medium, and light-colored contours enclose 68.3, 95.4, and 99.7\% of the probability, respectively. Dark dotted regions correspond to scenarios rejected by theoretical constraints.} 
\label{fig:gof}
\end{center}
\end{figure*}

Results using growth data are presented in Fig.~\ref{fig:gof} and Table~\ref{tab:results}, and are commented on in detail in Sect.~\ref{sec:disc}.

Growth data and cosmological distances provide consistent values for the $\bar c_i$s. The $\Omega_m^0$ best-fit value from growth data, $\Omega_m^0\approx 0.20$, is below that from the other probes, but is still compatible at the 1.5$\sigma$ level. 
This is the main difference between the two types of probes.

However, the use of growth data in cosmology deserves some comments. In our work, as in many others, different assumptions about the importance of non-linearities in structure formation are made in the theoretical predictions and in the experimental extraction of growth data from the measured matter power spectrum. 

As noted in Sect.~\ref{sec:pert}, our theoretical predictions are derived in the linear regime and using a quasi-static approximation. While \cite{bib:barreira} confirmed that the latter is valid in the Galileon model, using only the linear regime is restrictive. As an example, this may be the origin of the divergences in $G_{\mathrm{eff}}^{(\psi)}(z)/G_N$ that appear in some Galileon scenarios, as noted by \citet{bib:appleby2}. Going beyond the linear perturbation theory may change our predictions and thus could modify the result of our analysis.

To estimate this effect, we tried to identify at which scale non-linearities start to matter and checked whether this value is outside the range of scales taken into account in the growth-of-structure measurements.
As an example, WiggleZ measurements of $f\sigma_8$ are derived using a non-linear growth-of-structure model \citep{bib:jennings11} and encompass all scales  $k<0.3$~$h$Mpc$^{-1}$. In this model, the frontier between the linear and the non-linear regimes is $k\approx0.03$~$h$Mpc$^{-1}$. Other measurements in Table~\ref{tab:gof} include scales up to $k\approx0.2-0.4$~$h$Mpc$^{-1}$ as well.
On the other hand, there is no prediction in the Galileon model that goes beyond the linear regime. However, estimates of the scale at which non-linear effects appear exist in similar modified gravity models. Numerical simulations of the Chameleon screening effect for $f\left(R\right)$ theories show that non-linearity effects can be significant at scales $k\approx 0.05\ h$Mpc$^{-1}$ (see \cite{bib:brax12}, \cite{bib:jennings12} and \cite{bib:li}). Other simulations of the Vainshtein effect in the DGP model show that significant differences between the linear and non-linear regimes appear for scales $k>0.2\ h$Mpc$^{-1}$ \citep{bib:schmidt}. Unlike the DGP model, the Galileon model we considered does not contain a direct scalar-matter coupling $\sim \pi T^\mu_\mu$ that is usually considered as an essential ingredient of the Vainshtein effect. However, \cite{bib:babichev13} showed that even if the Galileon field is not directly coupled to matter,  the cosmological evolution of the Galileon field gives rise to an induced coupling of about 1, because of the Galileon-metric mixing. Therefore, the Vainshtein effect is expected to operate approximately at the same scales as in the DGP model in the model we considered.

This means the lack of non-linear effects in our perturbation equations, and hence in our predictions for $f\sigma_8$ in the Galileon model, is likely to have a significant impact on the constraints we derived from growth measurements, since the latter accounted partially for non-linear effects. 

\subsection{Full combined constraints}\label{sec:constraints}

\begin{figure*}[hbtp]
\begin{center}
\epsfig{figure=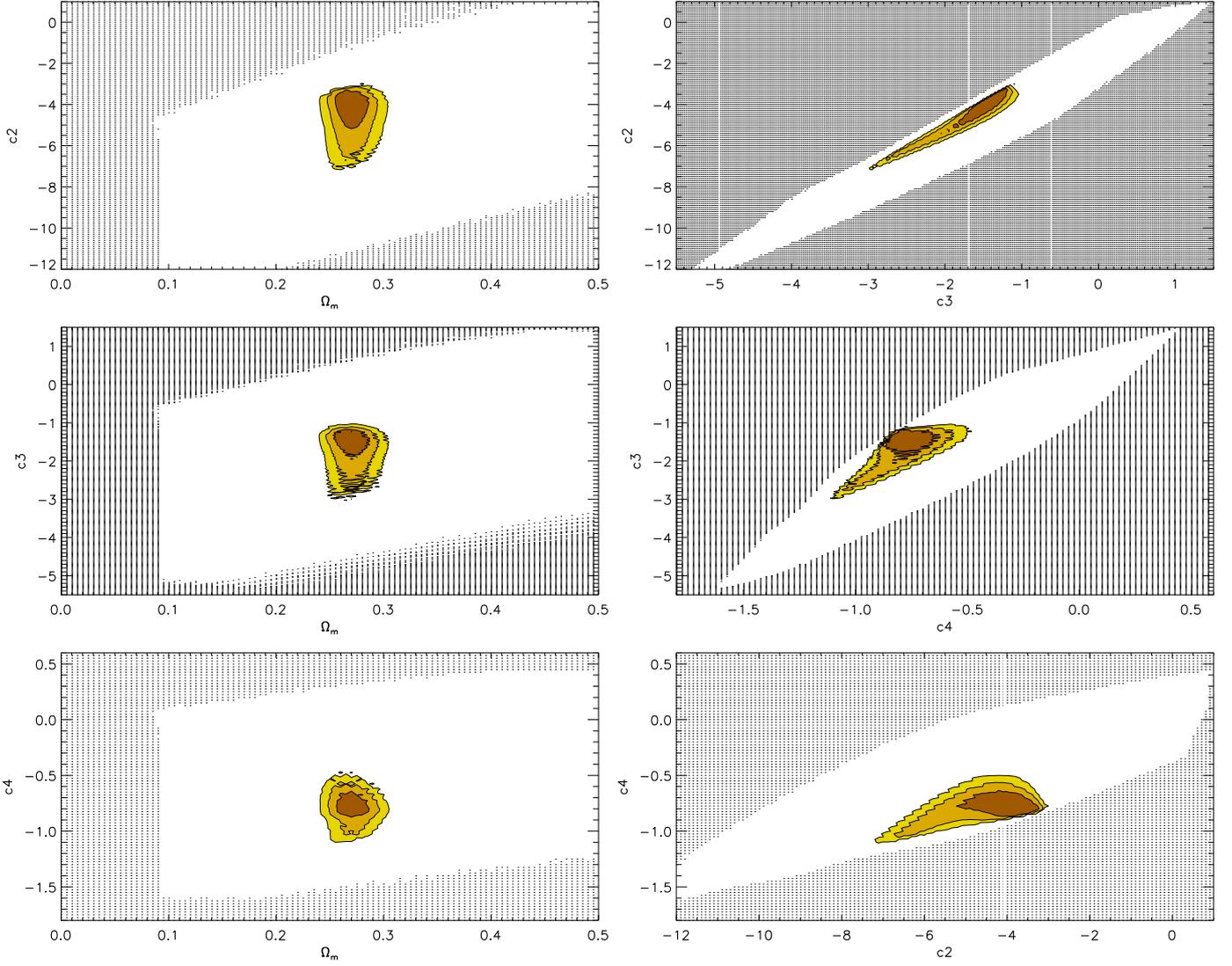, width=\textwidth} 
\caption[]{Combined constraints on the Galileon model from SNLS3, WMAP7+BAO+H0, and growth data. The filled dark, medium, and light-yellow contours enclose 68.3, 95.4, and 99.7\% of the probability, respectively. Dark dotted regions correspond to scenarios rejected by theoretical constraints.} 
\label{fig:combined}
\end{center}
\end{figure*}

Results from all data are presented in Fig.~\ref{fig:combined}. 
Table~\ref{tab:results} presents the best-fit values for the Galileon model parameters. The derived $\bar c_5$ value is\begin{equation}
\bar c_5^{\text{best-fit}}=-0.578^{+0.120}_{-0.219} .
\end{equation}
Note that negative values are preferred for the $\bar c_i$s at the 1$\sigma$ level. Moreover, the Galileon $h$ best-fit values are compatible with the \cite{bib:riess11} measurement.

We carried out an a posteriori check to identify which scenarios present a significant amount of early dark energy. At decoupling, $\Omega_{\pi}(z_*)>10 \% \Omega_r(z_*)$ only for viable scenarios with $\Omega_m^0>0.3,\ \bar c_2>-4,\ \bar c_3>-1$ and $\bar c_4>0$. This check can be made after comparing theory with data because only data can provide values for $h$ and $\Omega_b^0h^2$. For Galileon scenarios with $\Omega_m^0<0.3$, which is the region favoured by data, we found no significant early dark energy.

\begin{table*}[htb]
\caption[]{Galileon model best-fit values from different data samples}
\label{tab:results}
\begin{center}
\begin{tabular}{cccccccc} \hline \hline \\ [-1ex]
Probe & $\Omega_m^0$ & $\bar c_2$ & $\bar c_3$ & $\bar c_4$ & $h$ & $\Omega_b^0h^2$ & $ \chi^2$ \\  [1ex] \hline \\ [-1ex]
SNLS3 & $0.273^{+0.054}_{-0.042}$ & $-5.240^{+1.880}_{-2.802}$ & $-1.781^{+1.071}_{-1.426}$ & $-0.588^{+0.516}_{-0.348}$ & - & - & 420.1 \\ [1ex]  \hline \\ [-1ex]
Growth & $0.200^{+0.047}_{-0.044}$ & $-5.430^{+0.850}_{-1.563}$ & $-1.757^{+0.365}_{-1.251}$ & $-0.635^{+0.272}_{-0.179}$ & - & - & 19.83\\ [1ex]  \hline \\ [-1ex] 
BAO+WMAP7+H0  & $0.272^{+0.014}_{-0.009}$ & $-5.591^{+1.973}_{-2.655}$ & $-1.926^{+1.008}_{-1.407}$ & $-0.619^{+0.468}_{-0.335}$ & 0.713 & 0.0224 & 2.14\\ [1ex]  \hline \\ [-1ex] 
SNLS3+BAO+WMAP7+H0 & $0.272^{+0.014}_{-0.008}$ & $-5.565^{+1.959}_{-2.654}$ & $-1.917^{+1.001}_{-1.405}$ & $-0.619^{+0.468}_{-0.333}$ & 0.713 & 0.0224 & 423.1\\ [1ex]  \hline \\ [-1ex] 
SNLS3+BAO+WMAP7+H0+Growth & $0.271^{+0.013}_{-0.008}$ & $-4.352^{+0.518}_{-1.220}$ & $-1.597^{+0.203}_{-0.726}$ & $-0.771^{+0.098}_{-0.061}$ & 0.735 & 0.0220 & 450.4 \\ [1ex]  \hline 
\end{tabular}
\tablefoot{SNLS3 with systematics included, $\alpha$ and $\beta$ fixed to their marginalized value. $h$ and $\Omega_b^0h^2$ have been minimized so no error bars are provided.}
\end{center}

\caption[]{$\Lambda$CDM best-fit values from different data samples}
\label{tab:lcdm}
\begin{center}
\begin{tabular}{cccccc} \hline \hline \\ [-1ex]
Probe & $\Omega_m^0$ & $\Omega_\Lambda^0$ & $h$ & $\Omega_b^0h^2$ & $ \chi^2$ \\  [1ex] \hline \\ [-1ex]
SNLS3 & $0.178^{+0.100}_{-0.092}$ &  $0.664^{+0.170}_{-0.166}$ & - & - & 419.7 \\ [1ex]  \hline \\ [-1ex] 
Growth & $ 0.295^{+0.037}_{-0.031}$ & $0.646^{+0.067}_{-0.072}$ & - & - & 8.2   \\ [1ex]  \hline \\ [-1ex]
BAO+WMAP7+H0  & $0.288^{+0.014}_{-0.011}$ & $0.713^{+0.016}_{-0.014}$ & 0.691 & 0.0225 & 5.6  \\ [1ex]  \hline \\ [-1ex]
SNLS3+BAO+WMAP7+H0  & $0.283^{+0.013}_{-0.010}$ & $0.719^{+0.016}_{-0.013}$ & 0.692 & 0.0225 & 427.8  \\ [1ex]  \hline \\ [-1ex]
SNLS3+BAO+WMAP7+H0+Growth & $0.277^{+0.011}_{-0.009}$ & $0.725^{+0.015}_{-0.012}$ & 0.698 & 0.0225 & 440.2 \\ [1ex]  \hline
\end{tabular}
\tablefoot{SNLS3 with systematics included, $\alpha$ and $\beta$ fixed to their marginalized value. $h$ and $\Omega_b^0h^2$ have been minimized so no error bars are provided.}
\end{center}
\end{table*}

\subsection{Analysis of the best-fit scenario}\label{sec:bestfit}

\begin{figure*}[hbtp]
\begin{center}
\epsfig{figure=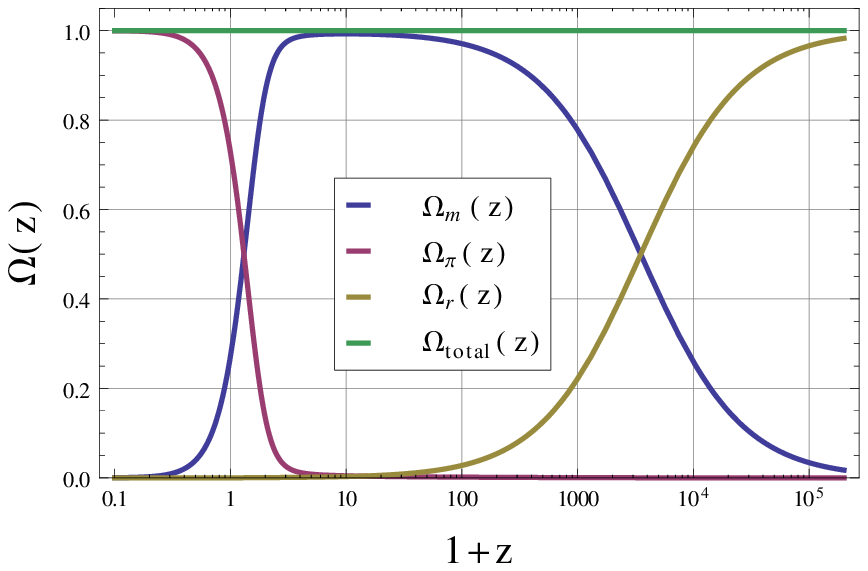, width=1.05\columnwidth} 
\epsfig{figure=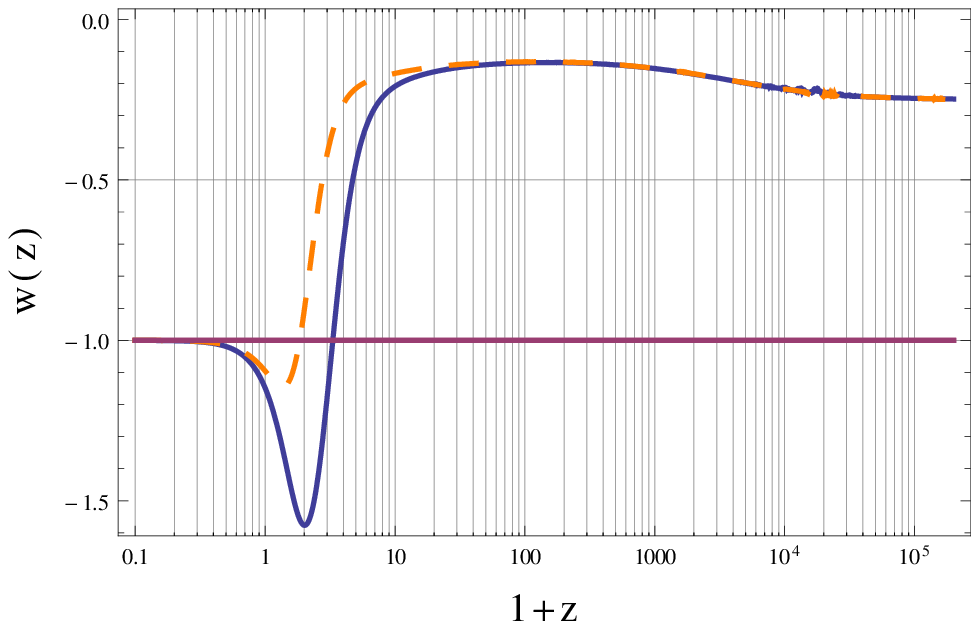, width=0.95\columnwidth} 
\caption[]{Evolution of the $\Omega_i(z)$ (left) and of $w(z)$ (right, solid curve) for the best-fit Galileon model from all data (last row of Table~\ref{tab:results}). As a comparison, the dashed orange line gives $w(z)$ for the best-fit scenario from SN data alone.} 
\label{fig:omega}
\end{center}
\end{figure*}

What does the best-fit scenario (derived from all data; the last line of table~\ref{tab:results}) look like? Because $\rho_\pi$ can be defined from the (00) Einstein equation, a Galileon pressure $P_\pi$ can be defined from the (ij) Einstein equation:
\begin{multline}\label{eq:ppi}
\frac{P_{\pi}}{H_0^2 M_P^2}= \frac{\bar c_2}{2}\h^2\x^2 + 2\bar c_3\h^3\x^2(\h \x)'\\  - \bar c_4\left[ \frac{9}{2}\h^6\x^4  + 12 \h^6\x^3\x' + 15 \h^5\x^4\h' \right]\\  + 3\bar c_5\h^7\x^4\left( 5\h \x' + 7\h' \x + 2\h \x \right) .
\end{multline}
Combining $\rho_\pi$ and $P_\pi$, an equation of state parameter $w_\pi(z)=P_\pi(z)/\rho_\pi(z)$ can be built for the Galileon "fluid". We can also construct an equation for $\Omega_\pi(z)$ using $\rho_\pi(z)=\Omega_\pi(z)H_0^2M_P^2/(3\bar H^2(z))$. The evolution of $w_\pi(z)$, $\Omega_\pi(z)$ and $G_{\mathrm{eff}}^{(\psi)}(z)/G_N$ for the Galileon best-fit scenario is shown in Figs.~\ref{fig:omega} and \ref{fig:geff}.

\subsubsection{Cosmic evolution}

The left panel of Fig.~\ref{fig:omega} shows that for the best-fit scenario, radiation, matter, and dark energy (here the Galileon) dominate alternatively during the history of the Universe, as in any standard cosmological model. These three epochs are also visible in the evolution of $w(z)$. Moreover, the best-fit scenario evolves in the future toward the de Sitter solution $w=-1$, which is an attractor of the Galileon model \citep{bib:felice2010}. In the region $0<z<1$, where SNe tightly constrain dark energy, $w(z)$ deviates significantly from -1, its $\Lambda$CDM value. Note that in the fit with SNe alone, the deviation is less pronounced, with an average value of -1.09 
in $0<z<1$, which is compatible with the fitted value of $w$ in constant $w$ dark energy models, as published in \cite{bib:conley}.

During matter domination, dark energy contributes about 0.4\% to the mass-energy budget at $z=10$. For comparison, in a standard $\Lambda$CDM model dark energy contributes only 0.2\% at this redshift (assuming a flat $\Lambda$CDM model with $\Omega_m^0=0.27$). In the same way, dark energy contributes 0.04\% at  $z_*$ in the Galileon best-fit scenario, whereas for $\Lambda$CDM $\Omega_\Lambda = 10^{-9}$ at $z_*$. In our best-fit Galileon scenario, dark energy is more present throughout the history of
the Universe than in the $\Lambda$CDM model, but is still negligible during the matter and radiation eras.




\begin{figure}[hbtp]
\begin{center}
\epsfig{figure=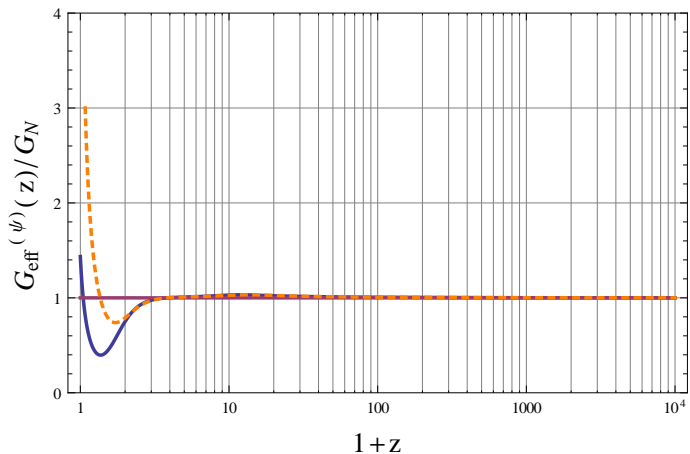, width=\columnwidth} 
\caption[]{Evolution of $G_{\mathrm{eff}}^{(\psi)}(z)/G_N$ for the best-fit scenario from growth data only (dashed orange line) and from all data (blue solid line).} \label{fig:geff}
\end{center}
\end{figure}

Figure~\ref{fig:geff} shows the evolution of $G_{\mathrm{eff}}^{(\psi)}(z)/G_N$ for the best-fit scenario and for the growth-data best-fit scenario. Both curves show deviations from 1 at redshifts around 0. Particularly, the divergence near the current epoch suggests that we should push the Galileon predictions for $f\sigma_8$ beyond the linear regime, as already advocated in Sect.~\ref{sec:growth}.

\subsubsection{Comparison with $\Lambda$CDM}

\begin{figure}[hbtp]
\begin{center}
\epsfig{figure=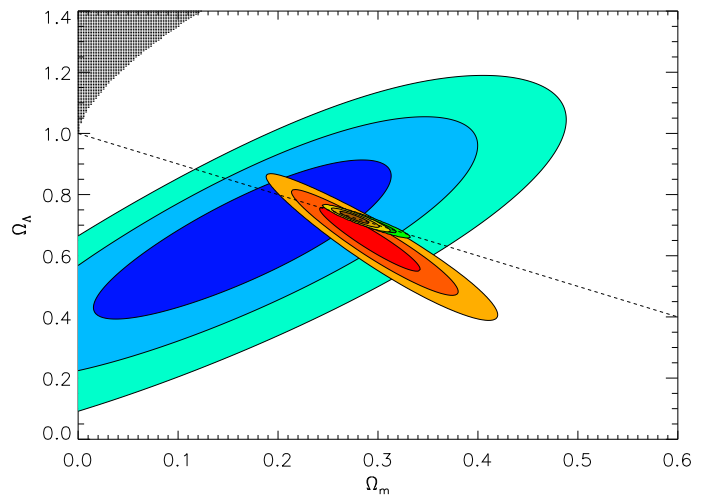, width=\columnwidth} 
\caption{Experimental constraints on the $\Lambda$CDM model from SNLS3 data (blue), growth  data (red), BAO+WMAP7+H0 data (green), and all data combined (yellow). The black dashed line indicates the flatness condition $\Omega_m+\Omega_\Lambda=1$.} 
\label{fig:lcdm}
\end{center}
\end{figure}

In Fig.~\ref{fig:lcdm} and Table~\ref{tab:lcdm}, best-fit values for the $\Lambda$CDM parameters are presented using the same analysis tools and observables. Interestingly, even in the $\Lambda$CDM model there is tension between growth data and other probes. The $\Omega_m^0$ best-fit value is similar in both models, but the $h$ value departs more from the $H_0$ \cite{bib:riess11} measurement. As far as the $\chi^2$s are concerned, SNe~Ia provide a good agreement with both models. CMB+BAO+H0 data are more compatible with the Galileon model, reflecting the better agreement on the $h$ minimized value. Yet growth-of-structure data agree better with the $\Lambda$CDM model. Finally, due to the poorer fit to growth data in the Galileon model, the difference in $\chi^2$ is $\Delta \chi^2 = 10.2$. This indicates that the Galileon model is slightly disfavored with respect to the $\Lambda$CDM model, despite having two extra free parameters.

Because we are comparing two models with a different number of parameters and complexity, other criteria than comparing $\chi^2$s can be helpful. A review of the selection model criterion is provided in \cite{bib:liddle07}. Because our study leads to the full computation of the likelihood functions, we can use precise criteria such as the Bayes factor (see \cite{bib:pdg}, \cite{bib:john}, \cite{bib:kass} and \cite{bib:liddle09}) or the deviance information criterion (DIC, see \cite{bib:spiegelhalter} and \cite{bib:kunz}). The Akaike information criterion (AIC) and the Bayesian information criterion (BIC) criteria used in \cite{bib:nesseris} are approximations of the first two using only the maximum likelihood and not the whole function. Hereafter we restrict the discussion to the DIC criterion. 

The DIC criterion is based on the computation of the deviance likelihoods $\mathrm{Dev}(\theta) = -2 \log p(D\mid \theta) + C$ (with C a constant not important for DIC evaluation). $p(\mathrm{D}\mid\theta)$ is the computed likelihood function $\mathcal{L}(\theta)$ of the model. An effective number of parameters $p_D=\overline{\mathrm{Dev}(\theta)}-\mathrm{Dev}(\overline{\theta})$ is derived with $\overline{\theta}$ the expectation values for $\theta$ and $\overline{\mathrm{Dev}(\theta)}$ the mean deviance likelihood value:
\begin{equation}
\overline{\mathrm{Dev}(\theta)} = -2\int d\theta p(\theta\mid \mathrm{D}) \log \mathcal{L}(\theta),
\end{equation}  
where $p(\theta \mid \mathrm{D})$ is the posterior probability density function for a vector $\theta$ of parameters of the tested model, knowing the data D:
\begin{equation}
p(\theta \mid \mathrm{D})=\frac{p(\mathrm{D}\mid\theta)\times \mathrm{prior}(\theta)}{p(\mathrm{D})}.
\end{equation}
$p(\mathrm{D})$, the probability to obtain the data D, is also called the marginal likelihood because it can be computed using the summation over all $\theta$s:
\begin{equation}
p(\mathrm{D})=\int d\theta\ p(\mathrm{D}\mid\theta)\times \mathrm{prior}(\theta)=\int d\theta \mathcal{L}(\theta)\times \mathrm{prior}(\theta).
\end{equation}
Note that if the priors are flat, $p(\theta \mid \mathrm{D})$ is just the likelihood function $\mathcal{L}(\theta)$ normalized to 1. In our case, $\mathrm{prior}(\theta)$ is a flat prior reflecting the theoretically allowed volume in the scanned parameter space. We checked that the DIC criterion is not sensitive to the exact definition of the prior, which makes it a robust tool.

Then $\mathrm{DIC}=\mathrm{Dev}(\overline{\theta})+2p_D=\overline{\mathrm{Dev}(\theta)}+p_D$. The model with the smallest DIC is favored by the data. In our study, we obtained $\mathrm{DIC}_{\mathrm{Galileon}}- \mathrm{DIC}_{\Lambda\mathrm{CDM}}=12.25>0$. Again, the Galileon model is slightly disfavored by data against the $\Lambda$CDM model. The DIC criterion just reflects the $\Delta\chi^2$ and does not penalize the Galileon model so much because of its higher number of free parameters. 

In the future, provided the tension between growth-of-structure data and distances does not increase after more precise measurements of the observables used in this paper are included, new observables will be necessary to distinguish between the two models. A promising way would be to exploit, e.g., the ISW effect as discussed in \cite{bib:kobayashi}.

\subsubsection{Comparison with FWCDM}

\begin{table*}[t]
\caption[]{FWCDM best-fit values from different data samples}
\label{tab:fwcdm}
\begin{center}
\begin{tabular}{cccccc} \hline \hline \\ [-1ex]
Probe & $\Omega_m^0$ & $w$ & $h$ & $\Omega_b^0h^2$ & $ \chi^2$ \\  [1ex] \hline \\ [-1ex]
SNLS3 & $0.183^{+0.095}_{-0.102}$ &  $-0.91^{+0.17 }_{-0.25}$ & - & - & 419.6 \\ [1ex]  \hline \\ [-1ex] 
Growth & $0.294^{+0.039}_{-0.030}$ & $-0.87^{+0.09}_{-0.08}$ & - & - & 7.9 \\ [1ex]  \hline \\ [-1ex]
BAO+WMAP7+H0  & $0.277^{+0.017}_{-0.012}$ & $-1.16^{+0.11}_{-0.11}$ & $0.718$ & $0.0222$ & 3.8  \\ [1ex]  \hline \\ [-1ex]
SNLS3+BAO+WMAP7+H0  & $0.279^{+0.015}_{-0.009}$ & $-1.12^{+0.08}_{-0.07}$ & $0.713$ & $0.0223$ & 425.1  \\ [1ex]  \hline \\ [-1ex]
SNLS3+BAO+WMAP7+H0+Growth & $0.280^{+0.014}_{-0.009}$ & $-0.99^{+0.05}_{-0.04}$ & $0.697$ & $0.0225$ & 440.2 \\ [1ex]  \hline
\end{tabular}
\tablefoot{SNLS3 with systematics included, $\alpha$ and $\beta$ fixed to their marginalized value. $h$ and $\Omega_b^0h^2$ have been minimized so no error bars are provided.}
\end{center}
\end{table*} 
 
\begin{figure}[hbtp]
\begin{center}
\epsfig{figure=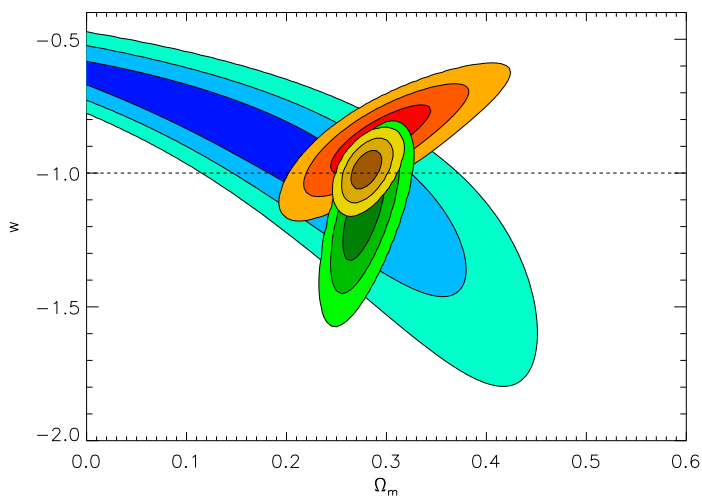, width=\columnwidth} 
\caption{Experimental constraints on the FWCDM model from SNLS3 data (blue), growth data (red), BAO+WMAP7+H0 data (green), and all data combined (yellow).} 
\label{fig:fwcdm}
\end{center}
\end{figure}

For consistency with our assumption about flatness, we also present a comparison with the effective FWCDM model, a model with a constant dark energy equation of state parameter $w$ in a flat Universe (see Table~\ref{tab:fwcdm} and Fig.~\ref{fig:fwcdm}). The data set points toward a value of $w$ below -1, which is consistent with the Galileon best-fit scenario (see Fig.~\ref{fig:omega}). 

However, the difference in $\chi^2$ is the same as for the $\Lambda$CDM model, $\Delta \chi^2 = 10.2$, and the DIC criterion gives $\mathrm{DIC}_{\mathrm{Galileon}}- \mathrm{DIC}_{\mathrm{FWCDM}}=12.16>0$. Here again, the Galileon model is not significantly disfavored.


\section{Discussion}\label{sec:disc}



In this section we compare our results with other recent publications on the same subject.

\citet{bib:appleby2} concluded that the uncoupled Galileon model is ruled out by current data since their best-fit yielded $\Delta \chi^2=31$ compared with the best-fit $\Lambda$CDM model. In addition, they obtained a long narrow region of degenerate scenarios with nearly the same likelihood. In our case, the best-fit has $\Delta \chi^2 =10.2$, we obtained enclosed  contours in all projections and a clear minimum.

Although we used the same expansion and perturbation equations as \cite{bib:appleby2}, there are differences between the two works. We used a parametrization of the model, which makes our study independent of initial conditions for $x$, while they set $x_i=x(z_i=10^6)$ by imposing a $\rho_\pi (z_i)$ which varied in their parameter scan. This requires one to solve a fifth-order polynomial equation in $x_i$ -- and hence one is forced to choose one of the five solutions -- or to assume one of the four terms $c_i \bar H^{2+2(i-2)} x^i$ is dominant in the (00) Einstein equation. In any case, this leads to a parameter space that is different than the one we explored.
Another difference arises from the theoretical constraints that are used to restrict the parameter space to viable scenarios only. Our set of theoretical constraints is larger because we also used tensorial constraints, which proved to be very powerful. This also leads to a different explored parameter space.
In \citet{bib:felice2010}, the rescaling of the Galileon parameters was performed with a de Sitter solution instead of using $x_0$, as in this paper. This led to relations fixing their "$\bar c_2$" and "$\bar c_3$" coefficients as a function of their "$\bar c_4$" and "$\bar c_5$" coefficients (denoted $\alpha$ and $\beta$ in their study), but required two initial conditions to compute the cosmological evolution. Those were also fitted using experimental data. With this parametrization and without growth constraints, \cite{bib:nesseris} found best-fit values for their "$\bar c_4$" and "$\bar c_5$" of the same sign and same order of magnitude as in our work, despite our different parametrizations. A second paper by \citet{bib:okada} included redshift space distortion measurements and ruled out the Galileon model at the 10$\sigma$ level. 

The first difference with respect to our work is the treatment of the initial conditions and the use of an extra theoretical constraint to avoid numerical instabilities during the transition from the matter era to the de Sitter epoch. This reduces the parameter space with respect to that explored in our work. 
As stated above, a better modeling of $G_{\mathrm{eff}}^{(\psi)}$ including non-linear effects should be conducted instead of discarding scenarios with such instabilities.

\begin{figure}[hbtp]
\begin{center}
\epsfig{figure=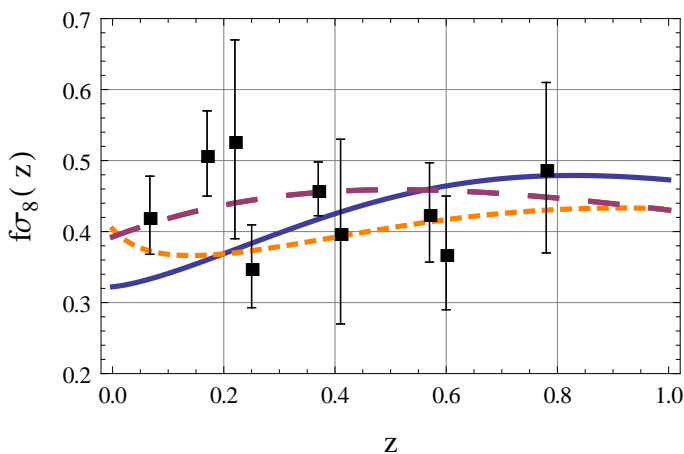, width=\columnwidth} 
\caption[]{$f\sigma_8(z)$ measurements from different surveys (6dFGRS, 2fFGRS, SDSS LRG, BOSS, and WiggleZ) compared with predictions for the $\Lambda$CDM model (with parameters of Table~\ref{tab:lcdm} - dashed purple line) Galileon scenarios. The solid blue line stands for the best-fit Galileon scenario using all data, whereas the orange dashed line stands for the best-fit Galileon scenario using growth data only.} \label{fig:fs8}
\end{center}
\end{figure}

\begin{figure}[hbtp]
\begin{center}
\epsfig{figure=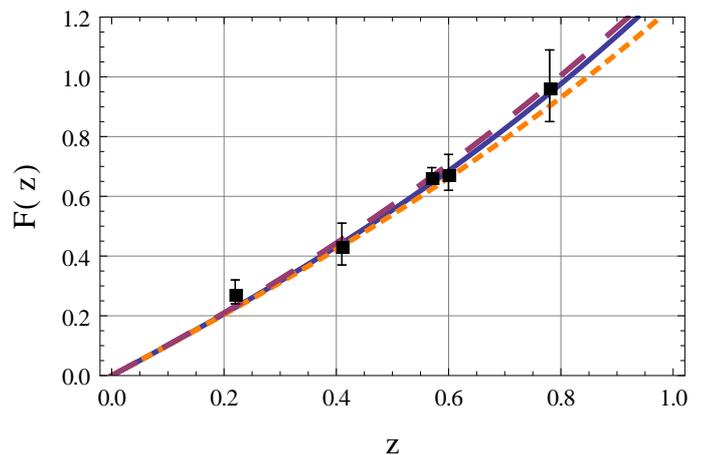, width=\columnwidth} 
\caption[]{$F(z)$ measurements from different surveys (BOSS and WiggleZ) compared with the prediction for the $\Lambda$CDM model (with parameters of Table~\ref{tab:lcdm} - dashed purple line) and for Galileon scenarios. The solid blue line stands for the best-fit Galileon scenario using all data, whereas the orange dashed line stands for the best-fit Galileon scenario using growth data only.} \label{fig:f}
\end{center}
\end{figure}

Second, \citet{bib:okada} used $f\sigma_8$ measurements not corrected for the Alcock-Paczynski effect. Moreover, to make their $f\sigma_8$ predictions in the Galileon model, \citet{bib:okada} set the normalization of $\sigma_8$ today to the WMAP7 $\sigma_8(z=0)$ measurement, which was obtained in a cosmological fit to the $\Lambda$CDM model. This normalization led to the following $\sigma_8(z)$ evolution:
\begin{equation}\label{eq:s83}
\sigma_8^{Gal}(z) = \sigma_8^{WMAP7}(z=0)\frac{D(z)}{D(0)}.
\end{equation}
This assumes that the Galileon theory predicts a matter power spectrum similar to that of $\Lambda$CDM at $z=0$, which is not guaranteed \citep{bib:barreira}. 
In contrast, we used the WMAP7 $\sigma_8$ measurement to set the normalization at decoupling $z\approx z_*$ (see Equation \ref{eq:s81}). Thus we took into account the different growth histories between the $\Lambda$CDM and the Galileon models (Equation \ref{eq:s82}, which is different from Equation \ref{eq:s83}). We can compare our best-fit scenarios for these two models with the $f\sigma_8$ and $F$ measurements. Figures~\ref{fig:fs8} and \ref{fig:f} show the result of this comparison. The agreement with the data is good in both models. In particular, the Galileon model does not exhibit a discrepancy as strong as was found in Fig. 3 of \cite{bib:okada}.

\section{Conclusion}\label{sec:concl}

We have confronted the uncoupled Galileon model with the most recent cosmological data.  We introduced a renormalization of the Galileon parameters by the derivative of the Galileon field normalized to the Planck mass to break some degeneracies inherent to the model. Theoretical conditions were added to restrict the analysis to viable scenarios only. This allowed us to break the parameter degeneracies that otherwise would have prevented us from obtaining enclosed probability contours. In particular, the conditions on the tensorial propagation mode of the perturbed metric proved to be very helpful. 

We used a grid search technique to explore the Galileon parameter space. Our data set encompassed the SNLS3 SN~Ia sample, WMAP7 $\left\lbrace l_a,R,z_* \right\rbrace$ constraints, BAO measurements, and growth data with the Alcock-Paczynski effect taken into account. We found $\left\lbrace \Omega_m^0,\bar c_2, \bar c_3, \bar c_4 \right\rbrace=\left\lbrace  0.271^{+0.013}_{-0.008},-4.352^{+0.518}_{-1.220},
-1.597^{+0.203}_{-0.726},-0.771^{+0.098}_{-0.061} \right\rbrace$. The final $\chi^2$ is slightly above that of the $\Lambda$CDM model due to a poorer fit to the growth data.

The best-fit Galileon scenario mimics a $\Lambda$CDM model with the three periods of radiation, matter, and dark energy domination, with an evolving dark energy equation of state parameter $w(z)$, and an effective gravitational coupling $G_{\mathrm{eff}}^{(\psi)}(z)$. Predictions for the latter are possible only in the linear regime, which may have an impact on our results derived from growth data because the latter were computed using a non-linear theory. A more precise theoretical and phenomenological study should be conducted to fairly compare the Galileon model with these data.

Our best-fit is more favorable to the Galileon model than other recent results. The main difference between our treatment and those works lies in the treatment of initial conditions.  We also tried to make as few assumptions and approximations as possible when computing observable quantities.  Finally, when using growth data, we took care to choose measurements that were derived in a model-independent way. In the future, a study considering precise predictions of the full power spectra as suggested by \citet{bib:barreira} would provide more stringent tests of the validity of the Galileon model.

\begin{acknowledgements} 

We thank Philippe Brax for introducing us to the Galileon theory, and Christos Charmoussis, C\'edric Deffayet, Jean-Baptiste Melin, and Marc Besan\c{c}on for fruitful discussions about the Galileon model. We also thank Chris Blake for useful advice on the use of the WiggleZ measurements.
The work of Eugeny Babichev was supported in part by grant FQXi-MGA-1209 from the Foundational Questions Institute.
\end{acknowledgements} 

\bibpunct{(}{)}{;}{a}{}{,} 

\appendix

\section{Instability of probability contours}\label{sec:appA}

Instead of absorbing the initial condition $x_0$ in the $c_i\rightarrow \bar c_i$ redefinition, we can be tempted to fix it using the (00)~Einstein equation at z=0 for each scenario:
\begin{equation}\label{eq:pi0}
1-\Omega_m^0 - \Omega_r^0 - \frac{1}{6}c_2 x_0^2 + 2c_3 x_0^3 - \frac{15}{2} c_4 x_0^4 + 7 c_5 x_0^5=0.
\end{equation}

\begin{figure}[hbtp]
\begin{center}
\epsfig{figure=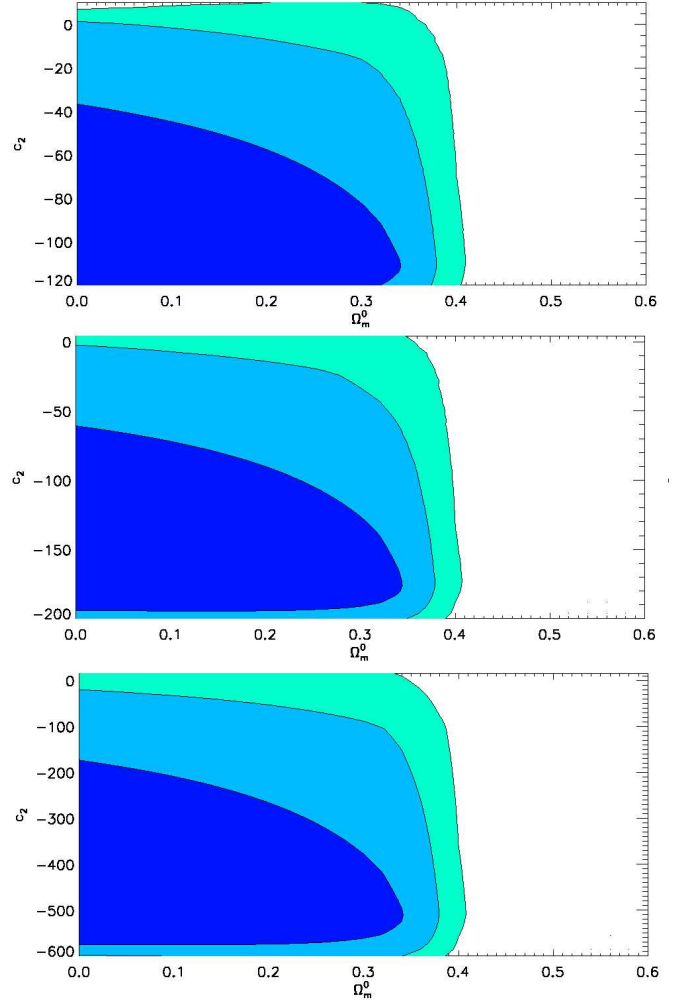, width=\columnwidth} 
\caption{Experimental constraints on the Galileon model from SNLS3 data for different ranges in $c_i$s using the method developed in Appendix A, with $\alpha$ and $\beta$ fixed to their $\Lambda$CDM best-fit values from \cite{bib:sullivan}. The four-dimensional likelihood $\mathcal{L}(\Omega_m^0,c_2, c_3, c_4)$ ($c_5$ fixed to 0 here) is marginalized over $c_3,c_4,\mathcal{M}_B^1$, $\mathcal{M}_B^2$ to visualize the $\Omega_m^0,c_2$ contour plots. The filled dark, medium, and light-blue contours enclose 68.3, 95.4, and 99.7\% of the probability, respectively.} 
\label{fig:instable}
\end{center}
\end{figure}

To find $x_0$, a fifth-order polynomial equation is to be solved, which can lead to five complex solutions. A reasonable choice is to keep only the scenarios that give a unique real solution. 

\begin{figure*}[hbtp]
\begin{center}
\epsfig{figure=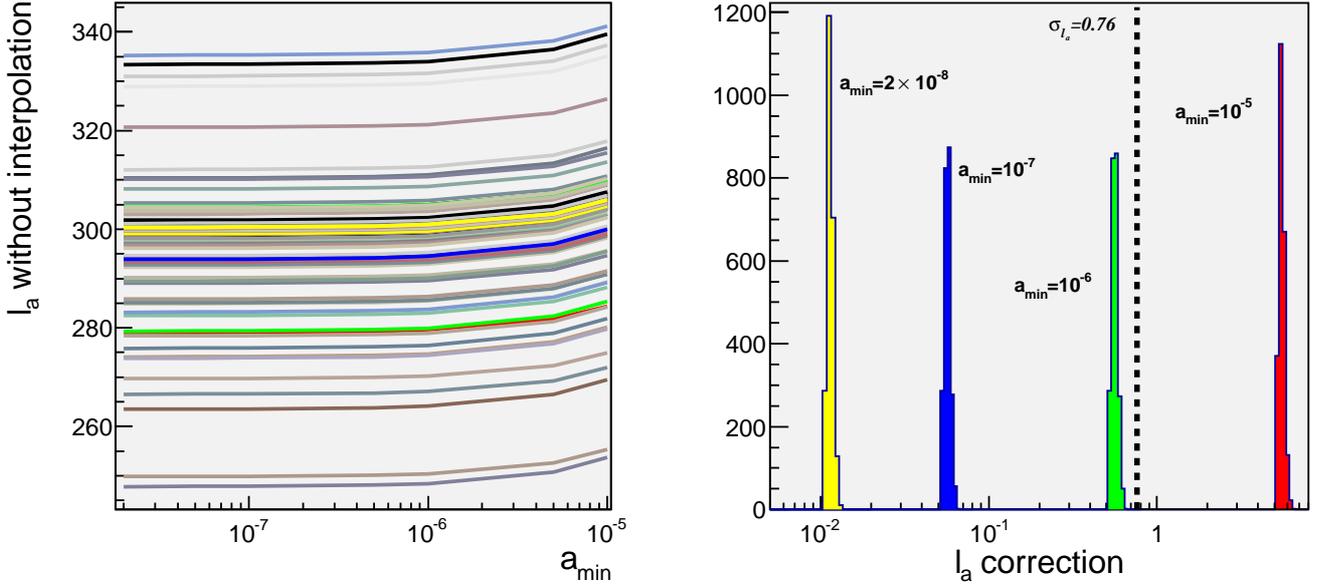, width=\textwidth} 
\caption{Left panel: evolution of $l_a$ with $a_{min}$ without the linear interpolation as described in the text for a subset of Galileon scenarios. Note that most scenarios approach the WMAP7 measurement $l_a \approx 300$. Right panel: correction to $l_a$ for different values of $a_{min}$ and for the same subset of scenarios as in the left panel. The dashed line is the value of $\sigma_{l_a}$, the WMAP7 measurement error on $l_a$.} 
\label{fig:laVsAmin}
\end{center}
\end{figure*}

The system of differential equations \ref{eq:dpi} and \ref{eq:dh} adopts an unusual behavior. Referring to Fig.~\ref{fig:instable}, the shape of the probability contours remains unchanged regardless of the limits of the scanned parameter space. In other words, the likelihood surface is invariant when the limits of the explored parameter space are proportionally changed. The model seems to exhibit a scale invariance allowing data to be fitted regardless of the boundaries of the explored parameter range. Moreover, we cannot obtain contours well enclosed in any explored parameter space: the likelihood surface has an infinite valley of minimum $\chi^2$ instead of a unique minimum.

Equation \ref{eq:pi0} shows that small $c_i$s produce a high $x_0$, and high $c_i$s a low $x_0$. Nevertheless, the theoretical constraints of Sect.~\ref{sec:theoconstraints} cannot favor or disfavor high $c_i$s or $x_0$ because they also contain this correspondence between the $c_i$s and $x_0$. Accordingly, for different sets of $c_i$s, identical cosmological scenarios are computed regardless of the scale of the $c_i$s: the important point is that these equivalent scenarios have the same $\Omega_{\pi}(z)$, whether this is due to high or small $c_i$s and as a consequence have the same $\h(z)$ evolution and then the same $\chi^2$.

Thus, a scale choice has to be made to fix the likelihood surface, but this choice has not to be arbitrary. A solution is provided in \ref{sec:inicond} by absorbing $x_0$ into new parameters $\bar c_i$s. This new parametrization absorbs a degree of freedom and allows us to use the (00) Einstein equation to fix $\bar c_5$. This may be the origin of the degeneracy in $\chi^2$ reported in Sect.~III of \cite{bib:appleby}.

\section{Approximation for $l_a$ computation}\label{sec:appB}

The computation of $l_a$ (see equation \ref{eq:la}) requires the evolution of the cosmological model from today to $a=0$ (see equation \ref{eq:rs}). In the Galileon context, the non-linear evolution equations require increasing precision and finer steps when approaching the limit $a \rightarrow 0$. In addition, it is physically questionable to extrapolate the Galileon model up to the very first instants of the Universe.

Therefore our iterative computation is stopped at a certain $a_{min}$ close to 0, without affecting significantly the final value of $l_a$. Let $a_{min-1}$ be the step before $a_{min}$ where the cosmological equations are computed, and $f(a)$ the integrand function of $r_s(z_*)$. Although the integral is stopped at $a=a_{min}$, we can compensate this approximation by a linear interpolation of the integral:

\begin{equation}
\begin{split}
r_s(z_*)\frac{H_0}{c}&=\int_0^\frac{1}{1+z_*}da\frac{\bar c_s(a)}{a^2\bar H(a)} = \int_0^\frac{1}{1+z_*}da f(a) \\ & \approx \int_{a_{min}}^\frac{1}{1+z_*}da f(a) + a_{min}f(a_{min}) \\ &\qquad\qquad - \frac{a_{min}^2}{2}\frac{f(a_{min-1})-f(a_{min})}{a_{min-1}-a_{min}}.
\end{split}
\end{equation}

In the left panel of Fig.~\ref{fig:laVsAmin}, we present the evolution of $l_a$ with  $a_{min}$ without the linear interpolation for a subset of Galileon scenarios. The smooth evolution with $a_{min}$ allows us to consider the linear interpolation as a reasonable assumption. Moreover, for $a_{min}\lesssim 10^{-6}$, the value of $l_a$ changes less than the WMAP7 measurement error $\sigma_{l_a}=0.76$, as shown in the right panel of Fig.~\ref{fig:laVsAmin}. 
Based on these results, we decide to use $a_{min}=10^{-7}$, which provides a correction on $l_a$ an order of magnitude below $\sigma_{l_a}$. 

\end{document}